\documentclass[11pt]{article} 
\newcommand{\doctype}{TECH}

\usepackage{fullpage}

\usepackage{amsmath}
\usepackage{amssymb}
\usepackage{amsfonts}
\usepackage{amsthm}

\usepackage{epsfig,graphicx}

\usepackage{epic,eepic}

\usepackage{epsf}
\usepackage{fancyhdr}
\usepackage{graphics}
\usepackage{amsfonts}
\usepackage{amsmath}
\usepackage{amssymb}
\usepackage{epic}
\usepackage{eepic}
\usepackage{eepicemu}
\usepackage{psfrag}

\usepackage{ifthen}

\ifthenelse{\equal{\doctype}{TECH}}
{

\setlength{\textwidth}{\paperwidth}
\addtolength{\textwidth}{-6cm}
\setlength{\textheight}{\paperheight}
\addtolength{\textheight}{-4cm}
\addtolength{\textheight}{-1.1\headheight}
\addtolength{\textheight}{-\headsep}
\addtolength{\textheight}{-\footskip}
\setlength{\oddsidemargin}{0.5cm}
\setlength{\evensidemargin}{0.5cm}

\setlength{\topmargin}{-1cm}
\setlength{\headheight}{0in}
\setlength{\headsep}{0in}
\setlength{\textheight}{9in}
\setlength{\oddsidemargin}{0in}
\setlength{\evensidemargin}{0in}
\setlength{\textwidth}{6.5in}
\addtolength{\headsep}{0.25in}
\marginparwidth 0.07 true in

}{}

\ifthenelse{\equal{\doctype}{AOS}}
{

\bibliographystyle{acmtrans-ims}

\startlocaldefs
}{}

\makeatletter

\theoremstyle{plain}

\newtheorem{theos}{Theorem}
\newtheorem{props}{Proposition}
\newtheorem{lems}{Lemma}
\newtheorem{cors}{Corollary}

\theoremstyle{definition}
\newtheorem{exas}{Example}
\newtheorem{algs}{Algorithm}
\newtheorem{asss}{Assumption}
\newtheorem{defns}{Definition}

\newcommand{\btheos}{\begin{theos}}
\newcommand{\etheos}{\end{theos}}
\newcommand{\bprops}{\begin{props}}
\newcommand{\eprops}{\end{props}}
\newcommand{\bdes}{\begin{defns}}
\newcommand{\edes}{\end{defns}}
\newcommand{\blems}{\begin{lems}}
\newcommand{\elems}{\end{lems}}
\newcommand{\bcors}{\begin{cors}}
\newcommand{\ecors}{\end{cors}}
\newcommand{\bexs}{\begin{exas}}
\newcommand{\eexs}{\end{exas}}
\newcommand{\balgs}{\begin{algs}}
\newcommand{\ealgs}{\end{algs}}
\newcommand{\bass}{\begin{asss}}
\newcommand{\eass}{\end{asss}}

\newcommand{\ignore}[1]{{}}
\newcommand{\half}{\ensuremath{\frac{1}{2}}}
\newcommand{\Paren}[1]{\ensuremath{\big({#1}\big)}}
\newcommand{\floor}[1]{\ensuremath{{\lfloor{#1}\rfloor}}}

\usepackage{amsfonts}
\usepackage{epsfig}
\usepackage{subfigure}
\usepackage{amsthm}
\usepackage{epic}
\usepackage{eepic}
\usepackage{epsf}
\usepackage{fancyhdr}
\usepackage{graphics}
\usepackage{amsfonts}
\usepackage{amsthm}
\usepackage{epic}
\usepackage{eepic}
\usepackage{eepicemu}
\RequirePackage[dvips]{hyperref}

\theoremstyle{plain}

\newtheorem{lemma}{Lemma}
\newtheorem{example}{Example}

\newcommand{\numobs}{\ensuremath{n}}

\newcommand{\graph}{\ensuremath{G}}
\newcommand{\vertex}{\ensuremath{V}}
\newcommand{\edge}{\ensuremath{E}}
\newcommand{\pdim}{\ensuremath{p}}

\newcommand{\kpar}{\ensuremath{k}}
\newcommand{\kdim}{\kpar}

\newcommand{\defn}{\ensuremath{: \, =}}
\newcommand{\Exs}{\ensuremath{\mathbb{E}}}
\newcommand{\prob}{\ensuremath{\mathbb{P}}}

\newcommand{\tmin}{\ensuremath{t^\ast}}

\newcommand{\cE}{{\cal E}}

\newcommand{\Gclass}{\ensuremath{\mathcal{G}}}

\newcommand{\real}{\ensuremath{\mathbb{R}}}

\newcommand{\mprob}{\ensuremath{\mathbb{P}}}

\long\def\comment#1{}

\newcommand{\degmax}{\ensuremath{d}}

\newcommand{\decoder}{\ensuremath{\phi}}
\newcommand{\decopt}{\ensuremath{\decoder^*}}
\newcommand{\dectwo}{\ensuremath{\decoder^\dagger}}

\newcommand{\Xsam}[1]{\ensuremath{X^{(#1)}}}

\newcommand{\Gcledge}{\ensuremath{\mathcal{G}_{\pdim, \kdim}}}
\newcommand{\Gcldeg}{\ensuremath{\mathcal{G}_{\pdim, \degmax}}}
\newcommand{\Neigh}{\ensuremath{\mathcal{N}}}

\newcommand{\eparam}{\ensuremath{\theta}}

\newcommand{\thetamin}{\ensuremath{\lambda}}

\newcommand{\speparam}[1]{\ensuremath{\eparam(#1)}}

\newcommand{\MyData}{\ensuremath{{\mathbf{X}_1^\numobs}}}
\newcommand{\StateSp}{\ensuremath{\mathcal{X}}}
\newcommand{\Ind}{\ensuremath{\mathbb{I}}}
\newcommand{\minvalfun}{\ensuremath{\lambda^*}}
\newcommand{\maxneighfun}{\ensuremath{\omega^*}}
\newcommand{\minval}{\ensuremath{\lambda}}
\newcommand{\maxneigh}{\ensuremath{\omega}}

\newcommand{\widgraph}[2]{\includegraphics[keepaspectratio,width=#1]{#2}}

\newcommand{\myorder}{\ensuremath{\mathcal{O}}}

\newcommand{\congen}{\ensuremath{c}}
\newcommand{\congentwo}{\ensuremath{c'}}
\newcommand{\contwo}{\ensuremath{c'}}

\newcommand{\kull}[2]{\ensuremath{D(#1 \, \| \, #2)}}
\newcommand{\bigkull}[2]{\ensuremath{D \big(#1 \, \| \, #2 \big)}}
\newcommand{\weirdkull}[2]{\ensuremath{J(#1 \, \| \, #2)}}

\newcommand{\symkull}[2]{\ensuremath{S(#1 \, \| \, #2)}}

\newcommand{\NumModel}{\ensuremath{M}}
\newcommand{\emodel}[1]{\ensuremath{\eparam^{(#1)}}}

\newcommand{\mint}{\ensuremath{m}}

\newcommand{\jstar}{\ensuremath{j^\ast}}

\newcommand{\myq}{\ensuremath{q_{st}}}

\newcommand{\myendproof}{\end{proof}}

\newcommand{\meanpar}{\ensuremath{\mu}}

\newcommand{\newgraph}{\ensuremath{H}}

\newlength{\widebarargwidth}
\newlength{\widebarargheight}
\newlength{\widebarargdepth}
\DeclareRobustCommand{\widebar}[1]{%
  \settowidth{\widebarargwidth}{\ensuremath{#1}}%
  \settoheight{\widebarargheight}{\ensuremath{#1}}%
  \settodepth{\widebarargdepth}{\ensuremath{#1}}%
  \addtolength{\widebarargwidth}{-0.3\widebarargheight}%
  \addtolength{\widebarargwidth}{-0.3\widebarargdepth}%
  \makebox[0pt][l]{\hspace{0.3\widebarargheight}%
    \hspace{0.3\widebarargdepth}%
    \addtolength{\widebarargheight}{0.3ex}%
    \rule[\widebarargheight]{0.95\widebarargwidth}{0.1ex}}%
  {#1}}

\newcommand{\graphbase}{\ensuremath{\widebar{\graph}}}

\newcommand{\likedecode}{\ensuremath{\decopt}}

\newcommand{\like}{\ensuremath{\ell}}
\newcommand{\matchnum}{\ensuremath{m}}

\newcommand{\newweird}[4]{\ensuremath{J^{#1}_{#2}(#3 \, \| \, #4)}}
\newcommand{\Aset}{\ensuremath{A}}

\newcommand{\Jcost}{\ensuremath{J}}

\newcommand{\estim}[1]{\ensuremath{\widehat{#1}}}

\newcommand{\Del}[2]{\ensuremath{\Delta(#1 ; #2)}}
\newcommand{\plweird}{\ensuremath{J}}

\newcommand{\unicon}{\ensuremath{c}}

\long\def\comment#1{}

\makeatletter
\def\@cite#1#2{[\if@tempswa #2 \fi #1]}
\makeatother

\makeatletter
\long\def\@makecaption#1#2{
        \vskip 0.8ex
        \setbox\@tempboxa\hbox{\small {\bf #1:} #2}
        \parindent 1.5em  
        \dimen0=\hsize
        \advance\dimen0 by -3em
        \ifdim \wd\@tempboxa >\dimen0
                \hbox to \hsize{
                        \parindent 0em
                        \hfil 
                        \parbox{\dimen0}{\def\baselinestretch{0.96}\small
                                {\bf #1.} #2
                                } 
                        \hfil}
        \else \hbox to \hsize{\hfil \box\@tempboxa \hfil}
        \fi
        }
\makeatother

\begin{document}

\long\def\mytitle{Information-theoretic limits of selecting binary
graphical models in high dimensions}

\long\def\abstext{
The problem of graphical model selection is to correctly estimate the
graph structure of a Markov random field given samples from the
underlying distribution. We analyze the information-theoretic
limitations of the problem of graph selection for binary Markov random
fields under high-dimensional scaling, in which the graph size $p$ and
the number of edges $k$, and/or the maximal node degree $d$ are allowed to
increase to infinity as a function of the sample size $n$. For
pairwise binary Markov random fields, we derive both necessary and
sufficient conditions for correct graph selection over the class
$\mathcal{G}_{p,k}$ of graphs on $p$ vertices with at most $k$ edges,
and over the class $\mathcal{G}_{p,d}$ of graphs on $p$ vertices with
maximum degree at most $d$.  For the class $\mathcal{G}_{p, k}$, we
establish the existence of constants $c$ and $c'$ such that if
$\numobs < c k \log p$, any method has error probability at least
$1/2$ uniformly over the family, and we demonstrate a graph decoder
that succeeds with high probability uniformly over the family for
sample sizes $\numobs > c' k^2 \log p$.  Similarly, for the class
$\mathcal{G}_{p,d}$, we exhibit constants $c$ and $c'$ such that for
$n < c d^2 \log p$, any method fails with probability at least $1/2$,
and we demonstrate a graph decoder that succeeds with high probability
for $n > c' d^3 \log p$.}

\ifthenelse{\equal{\doctype}{AOS}}
{
\begin{frontmatter}
\title{\mytitle}

\runtitle{Information-theoretic limits of graph selection}

\author{\fnms{Narayana}
\snm{Santhanam}\ead[label=e1]{prasadsn@gmail.com}\thanksref{t1}}
\affiliation{University of Hawaii} 

\author{\fnms{Martin J.}
\snm{Wainwright}\ead[label=e2]{wainwrig@stat.berkeley.edu}\thanksref{t1}}
\affiliation{University of California, Berkeley}

\thankstext{t1}{This work was partially supported by NSF grants
CAREER-0545862 and DMS-0528488 to MJW.  A preliminary
version of these results were presented in a talk at the International
Symposium on Information Theory in July 2008.}

\runauthor{Santhanam and Wainwright}

\begin{abstract}
\abstext
\end{abstract}

\begin{keyword}[class=AMS]
\kwd[Primary ]{62F12}
\kwd[; Secondary ]{68T99}
\end{keyword}

\begin{keyword}
\kwd{Graphical models}
\kwd{Markov random fields}
\kwd{Information theory}
\kwd{High-dimensional analysis}
\end{keyword}

\end{frontmatter}
}
{
}

\ifthenelse{\equal{\doctype}{TECH}}
{
\begin{center}

{\bf{\LARGE{\mytitle}}}

\vspace*{.2in}

\begin{tabular}{ccc}
Narayana Santhanam & \hspace*{.4in} & Martin J. Wainwright \\
Department of ECE & &  Departments of Statistics, and EECS \\
University of Hawaii & & UC Berkeley \\
Honolulu, HI  & &   Berkeley, CA  94720
\end{tabular}

\begin{abstract}
\abstext
\end{abstract}

\end{center}
}
{
}

\section{Introduction}

Markov random fields (also known as undirected graphical models)
provide a structured representation of the joint distributions of
families of random variables.  They are used in various application
domains, among them image analysis~\cite{Geman84,Besag86}, social
network analysis~\cite{Veg07,WasFau}, and computational
biology~\cite{DurbinEtal,KalBuh07,AhmSonXin08}. Any Markov random
field is associated with an underlying graph that describes
conditional independence properties associated with the joint
distribution of the random variables. The problem of \emph{graphical
model selection} is to recover this unknown graph using samples from
the distribution.

Given its relevance in many domains, the graph selection problem has
attracted a great deal of attention.  The naive approach of searching
exhaustively over the space of all graphs is computationally
intractable, since there $2^{\pdim \choose 2}$ distinct graphs over
$\pdim$ vertices.  If the underlying graph is known to be
tree-structured, then the graph selection problem can be reduced to a
maximum-weight spanning tree problem and solved in polynomial
time~\cite{Chow68}.  On the other hand, for general graphs with
cycles, the problem is known to be difficult in a complexity-theoretic
sense~\cite{chickering:95}.  Nonetheless, a variety of methods have
been proposed, including constraint-based
approaches~\cite{spirtes:00,KalBuh07}, thresholding
methods~\cite{BreMosSly08}, and $\ell_{1}$-based
relaxations~\cite{Meinshausen06,RavWaiLaf08,YuaLin07,FriedHasTib2007,Rot09}.
Other researchers~\cite{JiSey96,CsiTal06} have analyzed graph
selection methods based on penalized forms of pseudolikelihood.

Given a particular procedure for graph selection, a classical analysis
studies its behavior for a fixed graph as the sample size $\numobs$ is
increased.  In this paper, as with an evolving line of contemporary
statistical research, we address the graph selection problem in the
\emph{high-dimensional setting}, meaning that we allow the graph size
$\pdim$ as well as other structural parameters, such as the number of
edges $\kdim$ or the maximum vertex degree $\degmax$, to scale with
the sample size $\numobs$.  We note that a line of recent work has
established some high-dimensional consistency results for various
graph selection procedures, including methods based on
$\ell_1$-regularization for Gaussian
models~\cite{Meinshausen06,Ravetal08,Rot09}, $\ell_1$-regularization
for binary discrete Markov random fields~\cite{RavWaiLaf08},
thresholding methods for discrete models~\cite{BreMosSly08}, and
variants of the PC algorithm for directed graphical
models~\cite{KalBuh07}.  All of these methods are practically
appealing given their low-computational cost.

Of complementary interest---and the focus of the paper---are the
information-theoretic limitations of graphical model selection.  More
concretely, consider a graph $\graph = (\vertex, \edge)$, consisting
of a vertex set $\vertex$ with cardinality $\pdim$, and an edge set
$\edge \subset \vertex \times \vertex$.  In this paper, we consider
both the class $\Gcledge$ of all graphs with $|\edge| \le \kdim$ edges,
as well as the class $\Gcldeg$ all graphs with maximum vertex degree
$\degmax$.  Now suppose that we are allowed to collect $\numobs$
independent and identically distributed (i.i.d.) samples from a Markov
random field defined by some graph $\graph \in \Gcledge$ (or
$\Gcldeg$).  Remembering that the graph size $\pdim$ and structural
parameters $(\kdim, \degmax)$ are allowed to scale with the sample
size, we thereby obtain sequences of statistical inference problems,
indexed by the triplet $(\numobs, \pdim, \kdim)$ for the class
$\Gcledge$, and by the triplet $(\numobs, \pdim, \degmax)$ for the
class $\Gcldeg$.  The goal of this paper is to address questions of
the following type. First, under what scalings of the triplet
$(\numobs, \pdim, \kdim)$ (or correspondingly, the triplet $(\numobs,
\pdim, \degmax)$) is it possible to recover the correct graph with
high probability?  Conversely, under what scalings of these triplets
does any method fail most of the time?

Although our methods are somewhat more generally applicable, so as to
bring sharp focus to these issues, we limit the analysis of this paper
to the case of pairwise binary Markov random fields, also known as the
Ising model.  The Ising model is a classical model from statistical
physics~\cite{Ising25,Baxter}, where it is used to model physical
phenomena such as crystal structure and magnetism; more recently it
has been used in image analysis~\cite{Besag86,Geman84}, social network
modeling~\cite{BanGhaAsp08,Veg07}, and gene network
analysis~\cite{AhmSonXin08,SDM09:itw}.  

At a high level, then, the goal of this paper is to understand the
information-theoretic capacity of Ising model selection.\footnote{In
this paper, we assume that the data is drawn from some Ising model
from the class $\Gcledge$ and $\Gcldeg$, so that we study the
probability of recovering the exact model.  However, similar analysis
can be applied to the problem of find the best approximating
distribution using an Ising model from class $\Gcledge$ or
$\Gcledge$.}  Our perspective is not unrelated to a line of
statistical work in non-parametric
estimation~\cite{Hasminskii78,IbrHas81,Yu97,YanBar99}, in that we view
the observation process as a channel communicating information about
graphs to the statistician.  In contrast to non-parametric estimation,
the spaces of possible ``codewords'' are not function spaces but
rather classes of graphs. Accordingly, part of the analysis in this
paper involves developing ways in which to measure distances between
graphs, and to relate these distances to the Kullback-Leibler
divergence known to control error rates in statistical testing.  We
note that understanding of the graph selection capacity can be
practically useful in two different ways.  First, it can clarify when
computationally efficient algorithms achieve information-theoretic
limits, and hence are optimal up to constant factors.  Second, it can
reveal regimes in which the best known methods to date are
sub-optimal, thereby motivating the search for new and possibly better
methods.  Indeed, the analysis of this paper has consequences of both
types.

In this paper, we prove four main theorems, more specifically
necessary and sufficient conditions for the class $\Gcledge$ of
bounded edge cardinality models, and for the class $\Gcldeg$ of
bounded vertex degree models.  Proofs of the necessary conditions
(Theorems~\ref{ThmNecDeg} and~\ref{ThmNecEdge}) use indirect methods,
based on a version of Fano's lemma applied to carefully constructed
sub-families of graphs.  On the other hand, our proof of the
sufficient conditions (Theorems~\ref{ThmSufDeg} and~\ref{ThmSufEdge})
is based on direct analysis of explicit ``graph decoders''.  The
remainder of this paper is organized as follows.  We begin in
Section~\ref{SecBackground} with background on Markov random fields,
the classes of graphs considered in this paper, and a precise
statement of the graphical model selection problem.  In
Section~\ref{SecMain}, we state our main results and explore some of
their consequences.  Section~\ref{SecNecProof} is devoted to proofs of
the necessary conditions on the sample size (Theorems~\ref{ThmNecDeg}
and~\ref{ThmNecEdge}), whereas Section~\ref{SecSufProof} is devoted to
proofs of the sufficient conditions.  We conclude with a discussion in
Section~\ref{SecDiscuss}.

\noindent {\bf{Notation:}} For the convenience of the reader, we
summarize here notation to be used throughout the paper.  We use the
following standard notation for asymptotics: we write $f(n) =
\myorder(g(n))$ if $f(n) \leq \unicon g(n)$ for some constant $\unicon <
\infty$, and $f(n) = \Omega(g(n))$ if $f(n) \geq \unicon' g(n)$ for
some constant $\unicon' > 0$.  The notation $f(n) = \Theta(g(n))$
means that $f(n) = \myorder(g(n))$ and $f(n) = \Omega(g(n))$.


\section{Background and problem formulation}
\label{SecBackground}

We begin with some background on Markov random fields, and then
provide a precise formulation of the problem.

\subsection{Markov random fields and Ising models} 

An undirected graph $\graph = (\vertex, \edge)$ consists a collection
$\vertex = \{1, 2, \ldots, \pdim \}$ of vertices joined by a
collection $\edge \subset \vertex \times \vertex$ of
edges.\footnote{In this paper, we forbid self-loops in the graph,
meaning that $(s,s) \notin \edge$ for all $s \in \vertex$.}  The
neighborhood of any node $s \in \vertex$ is the subset $\Neigh(s)
\subset \vertex$
\begin{eqnarray}
\label{EqnDefnNeigh}
\Neigh(s) & \defn & \{t \in \vertex \, \mid \, (s,t) \in \edge \},
\end{eqnarray}
and the degree of vertex $s$ is given by $d_s \defn |\Neigh(s)|$,
corresponding to the cardinality of this neighbor set.  We use
$\degmax = \max_{s \in \vertex} d_s$ to denote the maximum vertex
degree, and $\kdim = |\edge|$ to denote the total number of edges.

A Markov random field is obtained by associating a random variable
$X_s$ to each vertex $s \in \vertex$, and then specifying a joint
distribution $\mprob$ over the random vector $(X_1, \ldots, X_\pdim)$
that respects the graph structure in a particular way.  In the special
case of the Ising model, each random variable $X_s$ takes values
$\{-1, +1\}$, and the the probability mass function has the form
\begin{eqnarray}
\label{EqnIsing}
\prob_\eparam(x_1, \ldots, x_\pdim) & = & \frac{1}{Z(\eparam)} \, \exp
\big \{\sum_{(s,t) \in \edge} \eparam_{st} x_s x_t\big \}
\end{eqnarray}
where $Z(\theta)$ is the normalization factor given by
\begin{eqnarray}
\label{EqnDefnZ}
Z(\eparam) & \defn & \log \Big[
\sum_{x \in \{-1,+1\}^\pdim} \exp \big \{\sum_{(s,t) \in \edge}
\eparam_{st} x_s x_t \big \} \Big].
\end{eqnarray}
To be clear, we view the parameter vector $\eparam$ as an element of
$\real^{{\pdim \choose 2}}$ with the understanding that $\eparam_{st}
= 0$ for all pairs $(s,t) \notin \edge$.  So as to emphasize the
graph-structured nature of the parameter $\eparam$, we often use the
notation $\speparam{\graph}$.  The edge weight $\eparam_{st}$
describes the conditional dependence between $X_s$ and $X_t$, given
fixed values for all vertices $X_u$, $u \neq s,t$.  In particular, a
little calculation shows that the conditional distribution takes the
form
\begin{eqnarray*}
\mprob_{\eparam} \big(x_s, x_t \, \mid \, x_{V \backslash \{s,t\}}
\big) & \propto & \exp \biggr( \eparam_{st} x_s x_t + \sum_{u \in
\Neigh(s) \backslash t} \eparam_{us} x_u x_s + \sum_{u \in \Neigh(t)
\backslash s} \eparam_{ut} x_u x_t \biggr).
\end{eqnarray*}

The Ising model~\eqref{EqnIsing} has its origins in statistical
physics~\cite{Ising25, Baxter}, where it used to model physical
phenomena such as crystal structure and magnetism; it is also has been
used as a simple model in image processing~\cite{Besag86,Geman84},
gene network analysis~\cite{AhmSonXin08,SDM09:itw}, and in modeling
social networks~\cite{BanGhaAsp08,Veg07}.  For instance, Banerjee et
al.~\cite{BanGhaAsp08} use this model to describe the voting behaviors
of $\pdim$ politicians, where $X_s$ represents whether politician $s$
voted for ($X_s = +1$) or against ($X_s = -1$) a particular bill.  In
this case, a positive edge weight $\eparam_{st} > 0$ would mean that
conditioned on the other politicians' votes, politician $s$ and $t$
are more likely to agree in their voting (i.e., $X_s = X_t$) than to
disagree ($X_s \neq X_t$), whereas a negative edge weight means that
they are more likely to disagree.

\subsection{Classes of graphical models}

In this paper, we consider two different classes of Ising
models~\eqref{EqnIsing}, depending on the condition that we impose on
the edge set $\edge$.  In particular, we consider the two classes of
graphs:
\begin{enumerate}
\item[(a)] the collection $\Gcldeg$ of graphs such that each vertex
  has degree at most $\degmax$ for some $\degmax \geq 1$, and
\item[(b)] the collection $\Gcledge$ of graphs $\graph$ with $|\edge|
  \leq \kdim$ edges for some $\kdim \geq 1$.
\end{enumerate}
In addition to the structural properties of the graphs, the difficulty
of graph selection also depends on properties of the vector of edge
weights $\speparam{\graph} \in \real^{{\pdim \choose 2}}$.  Naturally,
one important property is the minimum value over the edges.
Accordingly, we define the function
\begin{eqnarray}
\label{EqnMinValue}
\minvalfun(\speparam{\graph}) & \defn & \min_{(s,t) \in \edge}
|\eparam_{st}|.
\end{eqnarray}
The interpretation of the parameter $\thetamin$ is clear: as in
any signal detection problem, it is obviously difficult to detect an
interaction $\eparam_{st}$ if it is extremely close to zero.  

In contrast to classical signal detection problems, estimation of the
graphical structure turns out to be harder if the edge parameters
$\eparam_{st}$ are large, since the large value of edge parameters can
mask the presence of interactions on other edges.  The following
example illustrates this point:
\begin{example}
Consider the family $\Gcledge$ of graphs on $\pdim = 3$ with $\kdim =
2$ edges; note that there are a total of $3$ such graphs.  For each of
these three graphs, consider the parameter vector
\begin{eqnarray*}
\speparam{\graph} & = & \begin{bmatrix} \eparam & \eparam & 0
\end{bmatrix},
\end{eqnarray*}
where the single zero corresponds to the single distinct pair $s \neq
t$ \emph{not} in the graph's edge set, as illustrated in
Figure~\ref{FigSimple}.
\newcommand{\figw}{.18\textwidth}
\begin{figure}[h]
\begin{center}
\begin{tabular}{ccccc}
\psfrag{#a#}{$\eparam$}
\psfrag{#b#}{$\eparam$}
\psfrag{#c#}{$$}
\widgraph{\figw}{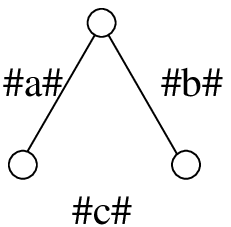} & \hspace*{.2in} &
\psfrag{#a#}{$\eparam$}
\psfrag{#b#}{$$}
\psfrag{#c#}{$\eparam$}
\widgraph{\figw}{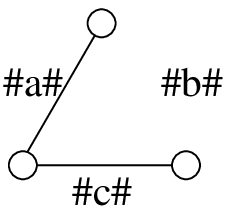} & \hspace*{.2in} &
\psfrag{#a#}{$$} \psfrag{#b#}{$\eparam$} \psfrag{#c#}{$\eparam$}
\widgraph{\figw}{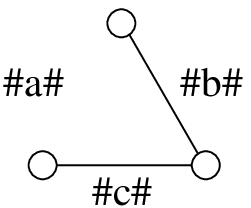} \\
(a) & & (b) & & (c)
\end{tabular}
\caption{Illustration of the family $\Gcledge$ for $\pdim = 3$ and
$\kdim = 2$; note that there are three distinct graphs $\graph$ with
$\pdim = 3$ vertices and $\kdim = 2$ edges.  Setting the edge
parameter $\speparam{\graph} = [ \eparam \; \eparam \; 0]$ induces a
family of three Markov random fields.  As the edge weight parameter
$\eparam$ increases, the associated distributions
$\mprob_{\speparam{\graph}}$ become arbitrarily difficult to
separate.}
\label{FigSimple}
\end{center}
\end{figure}
In the limiting case $\eparam = +\infty$, for any choice of graph with
two edges, the Ising model distribution enforces the ``hard-core''
constraint that $(X_1, X_2, X_3)$ must all be equal; that is, for any
graph $\graph$, the distribution $\mprob_{\speparam{\graph}}$ places
mass $1/2$ on the configuration $\begin{bmatrix} +1 & +1 & +1
\end{bmatrix}$ and mass $1/2$ on the configuration $\begin{bmatrix} -1
& -1 & -1
\end{bmatrix}$.  Of course, this hard-core limit is an extreme case,
in which the models are not actually identifiable.  Nonetheless, it
shows that if the edge weight $\eparam$ is finite but very large, the
models will not identical, but nonetheless will be extremely hard to
distinguish.  
\end{example}

Motivated by this example, we define the \emph{maximum neighborhood
weight}
\begin{eqnarray}
\label{EqnDefnMaxNeigh}
\maxneighfun(\speparam{\graph}) & \defn & \max_{s \in \vertex} \sum_{t
\in \Neigh(s)} |\theta_{st}|.
\end{eqnarray}
Our analysis shows that the number of samples $\numobs$ required to
distinguish graphs grows exponentially in this quantity.

In this paper, we study classes of Markov random fields that are
parameterized by a lower bound $\minval$ on the minimum edge weight,
and an upper bound $\maxneigh$ on the maximum neighborhood weight.

\bdes[{\bf{Classes of graphical models}}] 
\begin{enumerate}
\item[(a)] Given a pair $(\minval, \maxneigh)$ of positive numbers,
the set $\Gcldeg(\thetamin, \maxneigh)$ consists of all distributions
$\mprob_{\speparam{\graph}}$ of the form~\eqref{EqnIsing} such that
(i) the underlying graph $\graph = (\vertex, \edge)$ is a member of
the family $\Gcldeg$ of graphs on $\pdim$ vertices with vertex degree
at most $\degmax$; (ii) the parameter vector $\eparam =
\speparam{\graph}$ respects the structure of $\graph$, meaning that
$\eparam_{st} \neq 0$ only when $(s,t) \in \edge$, and (iii) the
minimum edge weight and maximum neighborhood satisfy the bounds
\begin{equation}
\label{eq:bounds}
\minvalfun(\speparam{\graph}) \, \geq \, \minval, \quad \mbox{and}
\quad \maxneighfun(\speparam{\graph}) \, \leq \, \maxneigh.
\end{equation}
\item[(b)] The set $\Gcledge(\thetamin, \maxneigh)$ is defined in an
analogous manner, with the graph $\graph$ belonging to the class
$\Gcledge$ of graphs with $\pdim$ vertices and $\kdim$ edges.
\end{enumerate}
\edes

We note that for any parameter vector $\speparam{\graph}$, we always
have the inequality
\begin{eqnarray}
\maxneighfun(\speparam{\graph}) & \geq & \max_{s \in \vertex}
|\Neigh(s)| \, \minvalfun(\speparam{\graph}),
\end{eqnarray}
so that the families $\Gcledge(\minval, \maxneigh)$ and
$\Gcldeg(\minval, \maxneigh)$ are only well-defined for suitable pairs
$(\minval, \maxneigh)$.

\subsection{Graph decoders and error criterion}

For a given graph class $\Gclass$ (either $\Gcldeg$ or $\Gcledge$) and
positive weights $(\minval, \maxneigh)$, suppose that nature chooses
some member $\mprob_{\speparam{\graph}}$ from the associated family
$\Gclass(\thetamin, \maxneigh)$ of Markov random fields.  Assume that
the statistician observes $\numobs$ samples $\MyData \defn \{
\Xsam{1}, \ldots, \Xsam{\numobs} \} $ drawn in an independent and
identically distributed (i.i.d.) manner from the distribution
$\mprob_{\speparam{\graph}}$.  Note that by definition of the Markov
random field, each sample $\Xsam{i}$ belongs to the discrete set
$\StateSp \defn \{-1, +1\}^\pdim$, so that the overall data set
$\MyData$ belongs to the Cartesian product space $\StateSp^\numobs$.

We assume that the goal of the statistician is to use the data
$\MyData$ to infer the underlying graph $\graph \in \Gclass$, which we
refer to as the problem of \emph{graphical model selection}.  More
precisely, we consider functions $\phi: \StateSp^\numobs \rightarrow
\Gclass$, which we refer to as graph decoders.  We measure the quality
of a given graph decoder $\phi$ using the 0-1 loss function
$\Ind[\phi(\MyData) \neq \graph]$, which takes value $1$ when
$\phi(\MyData) \neq \graph$ and takes the value $0$ otherwise, and we
define associated 0-1 risk
\begin{eqnarray*}
\mprob_{\speparam{\graph}} [\phi(\MyData) \neq \graph] & = &
\Exs_{\speparam{\graph}} \big[ \Ind[\phi(\MyData) \neq \graph] \big],
\end{eqnarray*}
corresponding to the probability of incorrect graph selection.  Here
the probability (and expectation) are taken with the respect the
product distribution of $\mprob_{\speparam{\graph}}$ over the $\numobs$
i.i.d. samples.

The main purpose of this paper is to study the scaling of the sample
sizes $\numobs$----more specifically, as a function of the graph size
$\pdim$, number of edges $\kdim$, maximum degree $\degmax$, minimum
edge weight $\minval$ and maximum neighborhood weight
$\maxneigh$---that are either sufficient for \emph{some} graph decoder
$\phi$ to output the correct graph with high probability, or
conversely, are necessary for \emph{any} graph decoder to output the
correct graph with probability larger than $1/2$.

We study two variants of the graph selection problem, depending on
whether the values of the edge weights $\eparam$ are known or unknown.
In the \emph{known edge weight} variant, \ignore{we assume that for
any fixed graph $\graph = (\vertex, \edge)$, the decoder knows the
numerical values of the parameters $\speparam{\graph}$.}  the task of
the decoder is to distinguish between graphs, where for any candidate
graph $\graph = (\vertex, \edge)$, the decoder knows the numerical
values of the parameters $\speparam{\graph}$.  (Recall that by
definition, $[\speparam{\graph}]_{uv} = 0$ for all $(u,v) \notin
\edge$, so that the additional information being provided are the
values $[\speparam{\graph}]_{st}$ for all $(s,t) \in \edge$.)  In the
\emph{unknown edge weight} variant, both the graph structure and the
numerical values of the edge weights are unknown.  Clearly, the
unknown edge variant is more difficult than the known edge variant.
We prove necessary conditions (lower bounds on sample size) for the
known edge variant, which are then also valid for the unknown variant.
In terms of sufficiency, we provide separate sets of conditions for
the known and unknown variants.

\section{Main results and some consequences}
\label{SecMain}

In this section, we state our main results and then discuss some of
their consequences.  We begin with statement and discussion of
necessary conditions in Section~\ref{SecNeg}, followed by sufficient
conditions in Section~\ref{SecSuff}.

\subsection{Necessary conditions}
\label{SecNeg}

We begin with stating some necessary conditions on the sample size
$\numobs$ that any decoder must satisfy for recovery over the families
$\Gcldeg$ and $\Gcledge$. Recall~\eqref{eq:bounds} for the definitions
of $\lambda$ and $\omega$ used in the theorems to follow.

\btheos[Necessary conditions for $\Gcldeg$]
\label{ThmNecDeg}
Consider the family $\Gcldeg(\minval, \maxneigh)$ of Markov random
fields for some $\maxneigh \geq 1$.  If the sample size is upper
bounded as
\begin{eqnarray}
\label{EqnNecDeg}
\numobs & \leq & \max \biggr \{ \frac{\log \pdim}{2 \minval
\tanh(\minval)}, \; \; \frac{\exp(\maxneigh/4) \degmax \minval \log
(\frac{\pdim \degmax}{4} -1)}{128 \exp(\frac{3 \minval}{2})}, \; \;
\frac{\degmax}{8} \log \frac{\pdim}{8 \degmax}, \biggr \},
\end{eqnarray}
then for any graph decoder $\decoder: \MyData \to \Gcldeg$, whether
given known edge weights or not,
\begin{eqnarray}
\max_{\speparam{\graph} \in \Gcldeg(\minval, \maxneigh)}
\mprob_{\speparam{\graph}} \big[ \decoder(\MyData) \neq \graph \big] &
\geq & \half.
\end{eqnarray}
\etheos
\noindent {\bf{Remarks:}} Let us make some comments regarding the
interpretation and consequences of Theorem~\ref{ThmNecDeg}. First,
suppose that both the maximum degree $\degmax$ and the minimum edge
weight $\minval$ remain bounded (i.e., do not increase with the
problem sequences).  In this case, the necessary
conditions~\eqref{EqnNecDeg} can be summarized more compactly as
requiring that for some constant $\congen$, a sample size $\numobs >
\frac{\congen \log \pdim}{\minval \tanh(\minval)}$ is required for
bounded degree graphs.  The observation of $\log \pdim$ scaling has
also been made in independent work~\cite{BreMosSly08}, although the
dependence on the signal-to-noise ratio $\minval$ given here is more
refined.  Indeed, note that if the minimum edge weight decreases to
zero as the sample size increases, then since $\minval \tanh(\minval)
= \myorder(\minval^2)$ for $\minval \rightarrow 0$, we conclude that a
sample size $\numobs > \frac{\congentwo \, \log \pdim}{\minval^2}$ is
required, for some constant $\congentwo$.

Some interesting phenomena arise in the case of growing maximum degree
$\degmax$.  Observe that in the family $\Gcldeg$, we necessarily have
$\maxneigh \geq \minval \degmax$.  Therefore, in the case of growing
maximum degree $\degmax \rightarrow +\infty$, if we wish the
bound~\eqref{EqnNecDeg} not to grow exponentially in $\degmax$, it is
necessary to impose the constraint $\minval =
\myorder(\frac{1}{\degmax})$.  But as observed previously, since
$\minval \tanh(\minval) = \myorder(\minval^2)$ as $\minval \rightarrow
0$, we obtain the following corollary of Theorem~\ref{ThmNecDeg}:
\bcors
\label{CorNecDeg}
For the family $\Gcldeg(\minval, \maxneigh)$ with increasing maximum
degree $\degmax$, there is a constant $\congen > 0$ such that in a
worst case sense, any method requires at least $\numobs > \congen \,
\max \{\degmax^2, \minval^{-2} \} \log \pdim$ samples to recover the
correct graph with probability at least $1/2$.
\ecors
We note that Ravikumar et al.~\cite{RavWaiLaf08} have shown that under
certain incoherence assumptions (roughly speaking, control on the
Fisher information matrix of the distributions
$\mprob_{\speparam{\graph}}$) and assuming that $\minval =
\Omega(\degmax^{-1})$, a computationally tractable method using
$\ell_1$-regularization can recover graphs over the family $\Gcldeg$
using $\numobs > \congentwo \, \degmax^3 \log \pdim$ samples, for some
constant $\congentwo$; consequently, Corollary~\ref{CorNecDeg} shows
concretely that this scaling is within a factor $\degmax$ of
information-theoretic bound.

We now turn to some analogous necessary conditions over the family
$\Gcledge$ of graphs on $\pdim$ vertices with at most $\kdim$ edges.
\btheos[Necessary conditions for $\Gcledge$]
\label{ThmNecEdge}
Consider the family $\Gcledge(\minval, \maxneigh)$ of Markov random
fields for some $\maxneigh \geq 1$.  If the sample size is upper
bounded as
\begin{eqnarray}
\label{EqnNecEdge}
\numobs & \leq & \max \biggr \{ \frac{\log \pdim}{2 \minval
  \tanh(\minval)}, \; \; \frac{ \exp(\frac{\maxneigh}{2}) \; \log
  (\kdim/8)}{64 \maxneigh \exp(\frac{5 \minval}{2}) \;
  \sinh(\minval)}, \biggr \},
\end{eqnarray}
then for any graph decoder $\decoder: \MyData \to \Gcledge$, whether
given known edge weights or not,
\begin{eqnarray}
\max_{\speparam{\graph} \in \Gcledge(\minval, \maxneigh)}
\mprob_{\speparam{\graph}} \big[ \decoder(\MyData) \neq \graph \big] &
\geq & \half.
\end{eqnarray}
\etheos

\noindent {\bf{Remarks:}}
Again, we make some comments about the consequences of
Theorem~\ref{ThmNecEdge}.  First, suppose that both the number of
edges $\kdim$ and the minimum edge weight $\minval$ remain bounded
(i.e., do not increase with the problem sequences).  In this case, the
necessary conditions~\eqref{EqnNecEdge} can be summarized more
compactly as requiring that for some constant $\congen$, a sample size
$\numobs > \frac{\congen \log \pdim}{\minval \tanh(\minval)}$ is
required for graphs with a constant number of edges.  Again, note that
if the minimum edge weight decreases to zero as the sample size
increases, then since $\minval \tanh(\minval) = \myorder(\minval^2)$
for $\minval \rightarrow 0$, we conclude that a sample size $\numobs >
\frac{\congentwo \, \log \pdim}{\minval^2}$ is required, for some
constant $\congentwo$.

The behavior is more subtle in the case of graph sequences in which
the number of edges $\kdim$ increases with the sample size.  As shown
in the proof of Theorem~\ref{ThmNecEdge}, it is possible to construct
a parameter vector $\speparam{\graph}$ over a graph $\graph$ with
$\kdim$ edges such that $\maxneighfun(\speparam{\graph}) \geq \minval
\lfloor \sqrt{\kdim} \rfloor$.  (More specifically, the construction
is based on forming a completely connected subgraph on $\lfloor
\sqrt{\kdim} \rfloor$ vertices, which has a total of ${\lfloor
\sqrt{\kdim} \rfloor \choose 2} \leq \kdim$ edges.)  Therefore, if we
wish to avoid the exponential growth from the term $\exp(\maxneigh)$,
we require that $\minval = \myorder(\kdim^{-1/2})$ as the graph size
increases.  Therefore, we obtain the following corollary of
Theorem~\ref{ThmNecEdge}:
\bcors
\label{CorNecEdge}
For the family $\Gcledge(\minval, \maxneigh)$ with increasing number
of edges $\kdim$, there is a constant $\congen > 0$ such that in a
worst case sense, any method requires at least $\numobs > \congen \,
\max \{ \kdim, \minval^{-2} \} \, \log \pdim$ samples to recover the
correct graph with probability at least $1/2$.
\ecors
To clarify a subtle point about comparing Theorems~\ref{ThmNecDeg}
and~\ref{ThmNecEdge}, consider a graph $\graph \in \Gcldeg$, say one
with homogeneous degree $\degmax$ at each node.  Note that such a
graph has a total of $\kdim = \degmax \pdim/2$ edges.  Consequently,
one might be misled into thinking Corollary~\ref{CorNecEdge} implies
that $\numobs > \congen \frac{\pdim \degmax}{2} \log \pdim$ samples
would be required in this case.  However, as shown in our development
of sufficient conditions for the class $\Gcldeg$ (see
Theorem~\ref{ThmSufDeg}), this is not true for sufficiently small
degrees $\degmax$.

To understand the difference, it should be remembered that our
necessary conditions are worst-case results, based on adversarial
choices from the graph families.  As mentioned, the necessary
conditions of Theorem~\ref{ThmNecEdge} and hence of
Corollary~\ref{CorNecEdge} are obtained by constructing a graph
$\graph$ that contains a completely connected graph,
$K_{\sqrt{\kdim}}$, with uniform degree $\sqrt{\kdim}$.  But
$K_{\sqrt{\kdim}}$ is not a member of $\Gcldeg$ unless $\degmax \geq
\sqrt{\kdim}$.  On the other hand, for the case when
$\degmax\geq\sqrt{\kdim}$, the necessary conditions of
Corollary~\ref{CorNecDeg} amount to $\numobs > \congen \kdim \log
\pdim$ samples being required, which matches the scaling given in
Corollary~\ref{CorNecEdge}.


\subsection{Sufficient conditions}
\label{SecSuff}

We now turn to stating and discussing sufficient conditions (lower
bounds on the sample size) for graph recovery over the families
$\Gcldeg$ and $\Gcledge$.  These conditions provide complementary
insight to the necessary conditions discussed so far.

\btheos[Sufficient conditions for $\Gcldeg$]
\label{ThmSufDeg}
(a) Suppose that for some $\delta \in (0,1)$, the sample size
$\numobs$ satisfies
\begin{eqnarray}
\label{EqnSufDegKnown}
\numobs & \geq & \frac{3 \big[3\exp(2 \maxneigh)+1\big]}{\sinh^2
 (\frac{\minval}{2})} \, \degmax \, \big \{ 3 \log \pdim + \log(2
 \degmax) + \log\frac{1}{\delta} \big \}.
\end{eqnarray}
Then there exists a graph decoder $\decopt: \MyData \rightarrow
\Gcldeg$ such that given known edge weights, the worst-case error
probability satisfies
\begin{eqnarray}
\max_{\speparam{\graph} \in \Gcldeg(\minval, \maxneigh)}
\mprob_{\speparam{\graph}} \big[ \decopt(\MyData) \neq \graph] & \leq
& \delta.
\end{eqnarray}

\noindent (b) In the case of unknown edge weights, suppose that the
sample size satisfies
\begin{eqnarray}
\label{EqnSufDegUnknown}
\numobs & > & \Big[\frac{\maxneigh \, \big(3 \exp(2 \maxneigh+1
\big)}{\sinh^{2}(\minval/4)}\Big]^2 \big \{ 16 \log \pdim + 4
\log(2/\delta) \big \}.
\end{eqnarray}
Then there exists a graph decoder $\dectwo: \MyData \rightarrow
\Gcldeg$ that that has worst-case error probability at most $\delta$.
\etheos

\noindent {\bf{Remarks:}} It is worthwhile comparing the sufficient
conditions provided by Theorem~\ref{ThmSufDeg} to the necessary
conditions from Theorem~\ref{ThmNecDeg}.  

First, consider the case of finite degree graphs. In this case, the
condition~\eqref{EqnSufDegKnown} reduces to the statement that for
some constant $\congen$, it suffices to have $\numobs > \congen \,
\minval^{-2} \log \pdim$ samples.  Comparing with the necessary
conditions (see the discussion following Theorem~\ref{ThmNecDeg}), we
see that for known edge weights and bounded degrees, the
information-theoretic capacity scales as $\minval^{-2} \, \log \pdim$.
For unknown edge weights, the conditions~\eqref{EqnSufDegUnknown}
provide a weaker guarantee, namely that $\numobs > \contwo \,
\minval^{-4} \, \log \pdim$ samples are required; we suspect that this
guarantee could be improved by a more careful analysis.

In the case of growing maximum graph degree $\degmax$, we note that
like the necessary conditions~\eqref{EqnNecDeg}, the sample size
specified by the sufficient conditions~\eqref{EqnSufDegKnown} scales
exponentially in the \mbox{parameter $\maxneigh$.}  If we wish not to
incur such exponential growth, we necessarily must have that $\minval
= \myorder(1/\degmax)$.    We thus obtain the following consequence
of Theorem~\ref{ThmSufDeg}:
\bcors
\label{CorSufDeg}
For the graph family $\Gcldeg(\minval, \maxneigh)$ with increasing
maximum degree, there exists a graph decoder that succeeds with high
probability using $\numobs > \congen_1 \, \max \{ \degmax^2, \,
\minval^{-2} \big \} \degmax \, \log \pdim$ samples.
\ecors
This corollary follows because the scaling $\minval =
\myorder(1/\degmax)$ implies that $\minval \rightarrow 0$ as $\degmax$
increases, and $\sinh(\minval/2) = \myorder(\minval)$ as $\minval
\rightarrow 0$. Note that in this regime, Corollary~\ref{CorNecDeg}
of Theorem~\ref{ThmNecDeg} showed that no method has error probability
below $1/2$ if $\numobs < \congen_2 \max \{ \degmax^2, \minval^{-2} \}
\, \log \pdim$, for some constant $\congen_2$.  Therefore, together
Theorems~\ref{ThmNecDeg} and~\ref{ThmSufDeg} provide upper and lower
bounds on the sample complexity of graph selection that are matching
to within a factor of $\degmax$.  We note that under the condition
$\minval \geq \frac{\congen_3}{\degmax}$, the results of Ravikumar et
al.~\cite{RavWaiLaf08} also guarantee correct recovery with high
probability for $\numobs > \congen_4 \degmax^3 \log \pdim$ using
$\ell_1$-regularized logistic regression; however, their method
requires additional (somewhat restrictive) incoherence assumptions
that are not imposed here.

Finally, we state sufficient conditions for the class $\Gcledge$ in the
case of known edge weights:
\btheos[Sufficient conditions for $\Gcledge$]
\label{ThmSufEdge}
(a) Suppose that for some $\delta \in (0,1)$, the sample size $\numobs$
satisfies
\begin{eqnarray}
\label{EqnSufEdge}
\numobs & > & \frac{3 \exp(2 \maxneigh)+1}{\sinh^2(\frac{\minval}{4})}
  \big((\kdim+1)\log \pdim +\log\frac{1}{\delta} \big).
\end{eqnarray}
Then for known edge weights, there exists a graph decoder $\decopt:
\MyData \rightarrow \Gcledge$ such that
\begin{eqnarray}
\max_{\speparam{\graph} \in \Gcledge(\minval, \maxneigh)}
\mprob_{\speparam{\graph}} \big[ \decopt(\MyData) \neq \graph] & \leq
& \delta.
\end{eqnarray}

\noindent (b) For unknown edge weights, there also exists a graph
decoder that succeeds under the condition~\eqref{EqnSufDegUnknown}.
\etheos
\noindent {\bf{Remarks:}} It is again interesting to compare
Theorem~\ref{ThmSufEdge} with the necessary conditions from
Theorem~\ref{ThmNecEdge}.  To begin, let the number
of edges $\kdim$ remain bounded.  In this case, for $\minval = o(1)$,
condition~\eqref{EqnSufEdge} states that for some constant $\congen$,
it suffices to have $\numobs > \frac{\congen \log \pdim}{\minval^2}$
samples, which matches (up to constant factors) the lower bound
implied by Theorem~\ref{ThmNecEdge}.  In the more general setting of
$\kdim \rightarrow +\infty$, we begin by noting that like in
Theorem~\ref{ThmNecEdge}, the sample size in Theorem~\ref{ThmSufEdge}
grows exponentially unless the parameter $\maxneigh$ stays controlled.
As with the discussion following Theorem~\ref{ThmNecEdge}, one
interesting scaling is to require that $\minval \asymp \kdim^{-1/2}$,
a choice which controls the worst-case construction that leads to
the factor $\exp(\maxneigh)$ in the proof of Theorem~\ref{ThmNecEdge}.
With this scaling, we have the following consequence:
\bcors
\label{CorSufEdge}
Suppose that the minimum value $\minval$ scales with the number
of edges $\kdim$ as $\minval \asymp \kdim^{-1/2}$.  Then in the
case of known edge weights, there exists a decoder that succeeds
with high probability using $\numobs > \congen \kdim^2 \log \pdim$
samples.
\ecors
\noindent Note that these sufficient conditions are within a factor of
$\kdim$ of the necessary conditions from Corollary~\ref{CorNecEdge},
which show that unless $\numobs > \contwo \max \{\kdim, \minval^{-2}
\} \, \log \pdim$, then any graph estimator fails at least half of the
time.


\section{Proofs of necessary conditions}
\label{SecNecProof}

In the following two sections, we provide the proofs of our main
theorems.  We begin by introducing some background on distances
between distributions, as well as some results on the cardinalities of
our model classes.  We then provide proofs of the necessary conditions
(Theorems~\ref{ThmNecDeg} and~\ref{ThmNecEdge}) in this section,
followed by the proofs of the sufficient conditions stated in
Theorems~\ref{ThmSufDeg} and~\ref{ThmSufEdge} in
Section~\ref{SecSufProof}.

\subsection{Preliminaries}

We begin with some preliminary definitions and results concerning
``distance'' measures between different models, and some estimates of
the cardinalities of different model classes.

\subsubsection{Distance measures}

In order to quantify the distinguishability of different models, we
begin by defining some useful ``distance'' measures.  Given two
parameters $\theta$ and $\theta'$ in $\real^{{\pdim \choose 2}}$, we
let $\kull{\theta}{\theta'}$ denote the Kullback-Leibler
divergence~\cite{Cover} between the two distributions $\mprob_\theta$
and $\mprob_{\theta'}$.  For the special case of the Ising model
distributions~\eqref{EqnIsing}, this Kullback-Leibler divergence takes
the form
\begin{eqnarray}
\label{EqnDefnKL}
\kull{\theta}{\theta'} & \defn & \sum_{x \in \{-1, +1\}^\pdim}
\mprob_\theta(x) \log \frac{\mprob_\theta(x)}{\mprob_{\theta'}(x)}.
\end{eqnarray}
Note that the Kullback-Leibler divergence is not symmetric in its
arguments (i.e., $\kull{\theta}{\theta'} \neq \kull{\theta}{\theta'}$
in general).

Our analysis also makes use of two other closely related divergence
measures, both of which are symmetric.  First, we define the
\emph{symmetrized Kullback-Leibler} divergence, defined in the natural
way via
\begin{eqnarray}
\label{EqnSymmetrizedKL}
\symkull{\eparam}{\eparam'} & \defn & \kull{\eparam}{\eparam'} +
\kull{\eparam'}{\eparam}.
\end{eqnarray}
Secondly, given two parameter vectors $\eparam$ and $\eparam'$, we may
consider the model $\mprob_{\frac{\eparam + \eparam'}{2}}$ specified
by their average.  In terms of this averaged model, we define another
type of divergence via
\begin{eqnarray}
\label{EqnWeirdKull}
\weirdkull{\eparam}{\eparam'} & \defn &
\kull{\frac{\eparam+\eparam'}{2}}{\eparam} +
\kull{\frac{\eparam+\eparam'}{2}}{\eparam'}.
\end{eqnarray}
Note that this divergence is also symmetric in its arguments.  A
straightforward calculation shows that this divergence measure can be
expressed in terms of the cumulant function~\eqref{EqnDefnZ}
associated with the Ising family as
\begin{eqnarray}
\label{EqnNewWeirdKull}
\weirdkull{\eparam}{\eparam'} & = & \log \frac{Z(\eparam)
Z(\eparam')}{Z^2(\frac{\eparam + \eparam'}{2})}.
\end{eqnarray}


Useful in our analysis are representations of these distance measures
in terms of the vector of \emph{mean parameters} $\meanpar(\eparam)
\in \real^{{\pdim \choose 2}}$, where element $\meanpar_{st}$ is given
by
\begin{eqnarray}
\label{EqnDefnMeanPar}
\meanpar_{st} & \defn & \Exs_{\eparam}[X_s X_t] \, = \, \sum_{x \in
\{-1, +1\}^\pdim} \mprob_\eparam[X] \, X_s X_t.
\end{eqnarray}
It is well-known from the theory of exponential
families~\cite{Brown86,WaiJor08} that there is a bijection between the
canonical parameters $\eparam$ and the mean parameters $\meanpar$.

Using this notation, a straightforward calculation shows that the
symmetrized Kullback-Leibler divergence between $\mprob_\eparam$ and
$\mprob_{\eparam'}$ is equal to
\begin{eqnarray}
\label{EqnNewSymKull}
\symkull{\eparam}{\eparam'} & = & \sum_{s,t \in \vertex, s \neq t}
\big(\eparam_{st} - \eparam'_{st} \big)\, \big(\meanpar_{st} -
\meanpar'_{st} \big),
\end{eqnarray}
where $\meanpar_{st}$ and $\meanpar'_{st}$ denote the edge-based mean
parameters under $\eparam$ and $\eparam'$ respectively.

\subsubsection{Cardinalities of graph classes}

In addition to these divergence measures, we require some estimates
of the cardinalities of the graph classes $\Gcldeg$ and $\Gcledge$,
as summarized in the following:

\begin{lemma} 
\label{LemCard}
\begin{enumerate}
\item[(a)] For $\kdim \leq {\pdim \choose 2}/2$, the
cardinality of $\Gcledge$ is bounded as
\begin{equation}
\label{EqnCardEdge}
{{\pdim \choose 2} \choose \kdim} \; \leq \; |\Gcledge| \; \leq \; 
\kdim  {{\pdim \choose 2} \choose \kdim},
\end{equation}
and hence $\log\big|\Gcledge \big | = \Theta (\kdim \log
\frac{\pdim}{\sqrt{\kdim}})$.
\item[(b)] For $\degmax \leq \frac{\pdim-1}{2}$, the cardinality of
$\Gcldeg$ is bounded as
\begin{equation}
\label{EqnCardDeg}
\biggr [\lfloor \frac{\pdim}{\degmax+1} \rfloor !
\biggr]^{\frac{\degmax(\degmax+1)}{2}} \; \leq \; |\Gcldeg| \; \leq \;
\frac{\pdim \degmax}{2} \, {{\pdim \choose 2 } \choose {\frac{\pdim
\degmax}{2}}}.
\end{equation}
and hence $\log\big|\Gcldeg \big | = \Theta \big(\pdim \degmax
  \log\frac{\pdim}{\degmax}\big)$.
\end{enumerate}
\end{lemma} 
\begin{proof}
(a) For the bounds~\eqref{EqnCardEdge} on $|\Gcledge|$, we observe
that there are ${{\pdim \choose 2} \choose \ell}$ graphs with exactly
$\ell$ edges, and that for $\kdim \leq {\pdim \choose 2}/2$, we have
${{\pdim \choose 2} \choose \ell} \leq {{\pdim \choose 2} \choose
\kdim}$ for all $\ell = 1,2, \ldots, \kdim$.

(b) Turning to the bounds~\eqref{EqnCardDeg} on $|\Gcldeg|$, observe
that every model in $\Gcldeg$ has at most $\frac{\pdim \degmax}{2}$
edges.  Note that $\degmax \leq \frac{\pdim-1}{2}$ ensures that
\[
\frac{\pdim \degmax}{2} \leq {\pdim \choose 2}/2.
\]  
Therefore, following the argument in part (a), we conclude that
$|\Gcldeg| \leq \frac{\pdim \degmax}{2} {{\pdim \choose 2} \frac{\pdim
\degmax}{2}}$ as claimed.  

In order to establish the lower bound~\eqref{EqnCardDeg}, we first
group the $\pdim$ vertices into $\degmax+1$ groups of size
$\floor{\frac{\pdim}{\degmax+1}}$, discarding any remaining
vertices. We consider a subset of $\Gcldeg$: graphs with maximum
degree $\degmax$ having the property that each component edge 
straddles vertices in two different groups.

To construct one such graph, we pick a permutation of
$\floor{\frac{\pdim}{\degmax+1}}$, and form an bijection from group
$1$ to group $2$ corresponding to the permutation. Similarly, we form
an bijection from group $1$ to $3$, and so on up until $\degmax+1$.
Note that use $\degmax$ permutations to complete this procedure, and
at the end of this round, every vertex in group 1 has degree $d$,
vertices in all other groups have degree 1.

Similarly, in the next round, we use $\degmax-1$ permutations to
connect group $2$ to groups $3$ through $\degmax+1$. In general, for
$i=1, \ldots, \degmax$, in round $i$, we use $d+1-i$ permutations to
connect group $i$ with groups $i+1, \ldots, \degmax+1$.  Each choice
of these permutations yields a distinct graph in
$\Gcldeg$.  Note that we use a total of
\begin{eqnarray*}
\sum_{i=1}^\degmax (\degmax +1 -i) & = & \sum_{\ell=1}^\degmax \ell \;
= \; \frac{\degmax \, (\degmax+1)}{2}
\end{eqnarray*}
permutations over $\lfloor \frac{\pdim}{\degmax+1} \rfloor$ elements,
from which the stated claim~\eqref{EqnCardDeg} follows.
\hfill \end{proof}


\subsubsection{Fano's lemma and variants}

We provide some background on Fano's lemma and its
variants needed in our arguments.  Consider a family of $\NumModel$
models indexed by the parameter vectors $\{\emodel{1}, \emodel{2},
\ldots, \emodel{\NumModel} \}$.  Suppose that we choose a model index
$k$ uniformly at random from $\{1, \ldots, \NumModel \}$, and than
sample a data set $\MyData$ of $\numobs$ samples drawn in an i.i.d.
manner according to a distribution $\mprob_{\emodel{k}}$.  In this
setting, Fano's lemma provides a lower bound on the probability of
error of any classification function $\decoder:
\StateSp^\numobs\rightarrow \{1, \ldots, \NumModel \}$, specified in
terms of the mutual information
\begin{eqnarray}
\label{EqnMutInfo}
I(\MyData; K) & = & H(\MyData) - H(\MyData \, \mid \, K)
\end{eqnarray}
between the data $\MyData$ and the random model index $K$.  We say
that a decoder $\decoder: \StateSp^\numobs \rightarrow \{1, \ldots, \NumModel
\}$ is unreliable over the family $\{\emodel{1}, \ldots,
\emodel{\NumModel} \}$ if
\begin{eqnarray}
\label{EqnUnreliable}
\max_{k = 1, \ldots, \NumModel} \mprob_{\emodel{k}}
\big[\decoder(\MyData) \neq k] & \geq & \frac{1}{2}.
\end{eqnarray}
We summarize Fano's inequality and a variant thereof in the following
lemma:

\begin{lemma}
\label{LemFano}
Any of the following upper bounds on the sample size imply that any
decoder $\decoder$ is unreliable over the family $\{ \emodel{1},
\ldots, \emodel{\NumModel} \}$:
\begin{enumerate}
\item[(a)] The sample size $\numobs$ is upper bounded as
\begin{eqnarray}
\label{EqnFano}
\numobs & < & \frac{\log(\NumModel/4)}{I(\MyData; K)}.
\end{eqnarray}
\item[(b)] The sample size $\numobs$ is upper bounded as
\begin{eqnarray}
\label{EqnFanoOne}
\numobs & < & \frac{\log (\NumModel/4)}{\frac{2}{\NumModel^2} \, \sum
\limits_{k=1}^\NumModel \sum \limits_{\ell = {k+1}}^\NumModel
\symkull{\emodel{k}}{\emodel{\ell}}}.
\end{eqnarray}
%
\end{enumerate}
\end{lemma}
These variants of Fano's inequality are standard and widely-used in
the non-parametric statistics literature
(e.g.,~\cite{Hasminskii78,IbrHas81,Yu97,YanBar99}); see Cover and
Thomas~\cite{Cover} for a statement and proof of the original Fano's
inequality.


\subsection{A key separation result}

In order to exploit the condition~\eqref{EqnFanoOne}, one needs to
construct families of models with relatively large cardinality
($\NumModel$ large) such that the models are all relatively close in
symmetrized Kullback-Leibler (KL) divergence.  Recalling the
definition~\eqref{EqnDefnMeanPar} of the mean parameters and the form
of the symmetrized KL divergence~\eqref{EqnNewSymKull}, we see that
control of the divergence between $\mprob_{\eparam}$ and
$\mprob_{\eparam'}$ can be achieved by ensuring that their respective
mean parameters $\meanpar_{st}$ $\meanpar_{st}$ stay relatively close
for all edges $(s,t)$ where the models differ.

In this section, we state and prove a key technical lemma that allows
us to control the mean parameters of a certain carefully constructed
class of models.  As shown in the proofs of Theorems~\ref{ThmNecDeg}
and~\ref{ThmNecEdge} to follow, this lemma allows us to gain good
control on the symmetrized Kullback-Leibler divergences between pairs
of models.  Our construction, which applies to any integer $\mint \geq
2$, is based on the following procedure.  We begin with the complete
graph on $\mint$ vertices, denoted by $K_\mint$.  We then form a set
of ${\mint \choose 2}$ graphs, each of which is a subgraph of
$K_\mint$, by \emph{removing} a particular edge.  Denoting by
$\graph^{st}$ the subgraph with edge $(s,t)$ removed, we define the
Ising model distribution $\mprob_{\speparam{\graph^{st}}}$ by setting
$[\speparam{\graph^{st}}]_{uv} = \minval$ for all edges $(u,v)$, and
$[\speparam{\graph^{st}}]_{st} = 0$.

The following lemma shows that the mean parameter $\meanpar_{st} =
\Exs_{\speparam{\graph^{st}}}[X_s X_t]$ approaches its maximum value
$1$ exponentially quickly in the parameter $\maxneigh = \minval
\mint$.

\blems
\label{LemKeySep}
Suppose that $\maxneigh = \minval \mint \geq 2$.  Then the likelihood
ratio on edge $(s,t)$ is lower bounded as
\begin{eqnarray}
\label{EqnInter}
\frac{\mprob_{\speparam{\graph^{st}}}[X_s X_t = +1]}
{\mprob_{\speparam{\graph^{st}}}[X_s X_t = -1]} \; = \;
\frac{\myq}{1-\myq} & \geq & \frac{\exp \big(\frac{\maxneigh}{2} -
\frac{3}{2} \minval \big)}{\mint+1}.
\end{eqnarray}
and moreover, the mean parameter over the pair $(s,t)$ is lower
bounded as
\begin{eqnarray}
\label{EqnMeanBound}
\Exs_{\speparam{\graph^{st}}}[X_s X_t] & \geq & 1-
\frac{2 \, (\mint+1)
\exp(\frac{3 \minval}{2})} {\exp(\frac{\maxneigh}{2}) + (\mint+1)
\exp(\frac{3 \minval}{2})}.
\end{eqnarray}
\elems
\begin{proof}
Let us introduce the convenient shorthand $\myq =
\mprob_{\speparam{\graph^{st}}}[X_s X_t = 1]$.  We begin by observing
that the bound~\eqref{EqnInter} implies the
bound~\eqref{EqnMeanBound}.  Indeed, suppose that
equation~\eqref{EqnInter} holds, or equivalently that $\myq \geq
\frac{b}{1+b}$ where $b = \frac{ \exp (\frac{\maxneigh}{2} -
\frac{3}{2} \minval)}{\mint+1}$.  Observing that
$\Exs_{\speparam{\graph^{st}}}[X_s X_t] = 2 \myq - 1$, we see that
$\frac{\myq}{1-\myq} \geq \frac{b}{1+b}$ implies that
\begin{eqnarray*}
\Exs_{\speparam{\graph^{st}}}[X_s X_t] & \geq & \frac{2 b}{1+b} - 1 \;
= \; 1 - \frac{2}{1 +b},
\end{eqnarray*}
from which equation~\eqref{EqnMeanBound} follows.

The remainder of our proof is devoted to proving the lower
bound~\eqref{EqnInter}.  Some calculation shows that
\begin{eqnarray}
\label{EqnRatio}
\frac{\myq}{1-\myq} & = & \frac{\sum_{j=0}^{\mint} {\mint \choose j}
\exp \big( \frac{\minval}{2} \big[(2j-\mint+1)^{2}-4\big] \big)}
{\sum_{j=0}^{\mint} {\mint \choose j}\exp\Paren{\frac{\lambda}{2} \big
[(2j-\mint)^{2}\big]}}.
\end{eqnarray}
We lower bound the ratio~\eqref{EqnRatio} by choosing one of largest
terms in the denominator.  It can be shown that for $\minval \mint
\geq 2$, the largest terms always lie in the range $j > 3 \mint/4$ and
$j < \mint/4$.  Accordingly, we may choose a maximizing point $\jstar
> 3 \mint/4$.  Since all the terms in the numerator are non-negative,
we have
\begin{eqnarray*}
\frac{\mprob_{\speparam{\graph^{st}}}[X_s X_t = +1]}
{\mprob_{\speparam{\graph^{st}}}[X_s X_t = -1]} & \geq & \frac{ {\mint
\choose \jstar} \exp \Paren{\frac{\minval}{2}
\big[(2\jstar-\mint+1)^2-4 \big]}} {(\mint+1) {\mint \choose \jstar}
\exp \Paren{\frac{\minval}{2} \big[(2 \jstar - \mint)^{2}\big]}} \\
& = & \frac{\exp\Paren{\frac{\minval}{2} \big[4 \jstar - 2\mint- 3
\big]}}{\mint+1}\\
& \geq & \frac{\exp\Paren{\frac{\minval}{2}\bigl[ \mint - 3
\big]}}{\mint+1}\\
& = & \frac{\exp\Paren{\frac{\maxneigh}{2} - \frac{3}{2}
\minval}}{\mint+1},
\end{eqnarray*}
which completes the proof of the bound~\eqref{EqnInter}. \\

\hfill \end{proof}

\subsection{Proof of Theorem~\ref{ThmNecDeg}}

We begin with necessary conditions for the bounded degree family
$\Gcldeg$.  The proof is based on applying Fano's inequality to three
ensembles of graphical models, each contained within the family
$\Gcldeg(\minval, \maxneigh)$.  \\

\noindent {\bf{Ensemble A:}} In this ensemble, we consider the set of
${\pdim \choose 2}$ graphs, each of which contains a single edge.  For
each such graph---say the one containing edge $(s,t)$, which we denote
by $\newgraph_{st}$---we set $[\speparam{\newgraph_{st}}]_{st} =
\minval$, and all other entries equal to zero.  Clearly, the resulting
Markov random fields $\mprob_{\speparam{\newgraph}}$ all belong to the
family $\Gcldeg(\minval, \maxneigh)$.  (Note that by definition, we
must have $\maxneigh \geq \minval$ for the family to be non-empty.)

Let us compute the symmetrized Kullback-Leibler divergence between the
MRFs indexed by $\speparam{\graph_{st}}$ and $\speparam{\graph_{uv}}$.
Using the representation~\eqref{EqnNewSymKull}, we have
\begin{eqnarray*}
\symkull{\speparam{\newgraph_{st}}}{\speparam{\newgraph_{uv}}} & = & \minval
\, \biggr \{ \big(\Exs_{\speparam{\newgraph_{st}}}[X_s X_t] -
\Exs_{\speparam{\newgraph_{uv}}}[X_s X_t] \big) -
\big(\Exs_{\speparam{\newgraph_{st}}}[X_u X_v] -
\Exs_{\speparam{\newgraph_{uv}}}[X_u X_v] \big) \biggr \} \\
& = & 2 \minval \Exs_{\speparam{\newgraph_{st}}}[X_s X_t],
\end{eqnarray*}
since $\Exs_{\speparam{\newgraph_{st}}}[X_u X_v] = 0$ for all $(u,v)
\neq (s,t)$, and $\Exs_{\speparam{\newgraph_{uv}}}[X_u X_v] =
\Exs_{\speparam{\newgraph_{st}}}[X_s X_t]$.  Finally, by definition of
the distribution $\mprob_{\speparam{\newgraph_{st}}}$, we have
\begin{eqnarray*}
\Exs_{\speparam{\newgraph_{st}}}[X_s X_t] & = & \frac{\exp(\minval) -
\exp(-\minval)}{\exp(\minval) + \exp(-\minval)} \; = \;
\tanh(\minval),
\end{eqnarray*}
so that we conclude that the symmetrized Kullback-Leibler divergence
is equal to $2 \minval \tanh(\minval)$ for each pair.

Using the bound~\eqref{EqnFanoOne} from Lemma~\ref{LemFano} with
$\NumModel = {\pdim \choose 2}$, we conclude that the graph
recovery is unreliable (i.e., has error probability above $1/2$)
if the sample size is upper bounded as
\begin{eqnarray}
\label{EqnBoundOne}
\numobs & < & \frac{\log ({\pdim \choose 2}/4)}{\minval
\tanh(\minval)}.
\end{eqnarray}

\noindent {\bf{Ensemble B:}} In order to form this graph ensemble, we
begin with a grouping of the $\pdim$ vertices into
$\floor{\frac{\pdim}{\degmax+1}}$ groups, each with $\degmax+1$
vertices.  We then consider the graph $\graphbase$ obtained by fully
connecting each subset of $\degmax+1$ vertices.  More explicitly,
$\graphbase$ is a graph that contains
$\floor{\frac{\pdim}{\degmax+1}}$ cliques of size $\degmax + 1$.
Using this base graph, we form a collection of graphs by beginning
with $\graphbase$, and then removing a single edge $(u,v)$.  We denote
the resulting graph by $\graph^{uv}$.  Note that if $\pdim \geq 2
(\degmax + 1)$, then we can form
\begin{eqnarray*}
\floor{\frac{\pdim}{\degmax+1}} {\degmax+1 \choose 2} & \geq &
\frac{\pdim \degmax}{4}
\end{eqnarray*}
such graphs.  For each graph $\graph^{uv}$, we form an associated
Markov random field $\mprob_{\speparam{\graph^{uv}}}$ by setting
$[\speparam{\graph^{uv}}]_{ab} = \minval > 0$ for all $(a,b)$ in the
edge set of $\graph^{uv}$, and setting the parameter to zero
otherwise.

A central component of the argument is the following bound on the
symmetrized Kullback-Leibler divergence between these distributions
\blems
\label{LemKeyDeg}
For all distinct pairs of models $\speparam{\graph^{st}} \neq
\speparam{\graph^{uv}}$ in ensemble B and for all $\minval \geq
1/\degmax$, the symmetrized Kullback-Leibler divergence is upper
bounded as
\begin{eqnarray*}
\symkull{\speparam{\graph^{st}}}{\speparam{\graph^{uv}}} & \leq &
\frac{8 \minval \, \degmax \exp(\frac{3 \minval}{2})}
{\exp(\frac{\minval \degmax}{2})}.
\end{eqnarray*}
\elems
\begin{proof}
Note that any pair of distinct parameter vectors
$\speparam{\graph^{st}} \neq \speparam{\graph^{uv}}$ differ in exactly
two edges.  Consequently, by the representation~\eqref{EqnNewSymKull},
and the definition of the parameter vectors,
\begin{eqnarray*}
  \symkull{\speparam{\graph^{st}}}{\speparam{\graph^{uv}}} & = & 
\minval
\big( \Exs_{\speparam{\graph^{uv}}}[X_s X_t] -
\Exs_{\speparam{\graph^{st}}}[X_s X_t] \big) + \minval \big(
\Exs_{\speparam{\graph^{st}}}[X_u X_v] -
\Exs_{\speparam{\graph^{uv}}}[X_u X_v] \big) \\
& \leq & \minval \big( 1 -\Exs_{\speparam{\graph^{st}}}[X_s X_t] \big)
+ \minval \big( 1- \Exs_{\speparam{\graph^{uv}}}[X_u X_v] \big),
\end{eqnarray*}
where the inequality uses the fact that $\minval > 0$, and the
edge-based mean parameters are upper bounded by $1$.

Since the model $\mprob_{\speparam{\graph^{st}}}$ factors as a product
of separate distributions over the $\lfloor \frac{\pdim}{\degmax+1}
\rfloor$ cliques, we can now apply the separation
result~\eqref{EqnMeanBound} from Lemma~\ref{LemKeySep} with $\mint =
\degmax+1$ to conclude that
\begin{eqnarray*}
  \symkull{\speparam{\graph^{st}}}{\speparam{\graph^{uv}}} & \leq & 2
\minval \, \frac{2 \, (\degmax+2) \exp(\frac{3 \minval}{2})}
{\exp(\frac{\minval (\degmax+1)}{2}) + (\degmax+2) \exp(\frac{3
\minval}{2})} \\
& \leq & \frac{8 \: \minval \: \degmax \: \exp(\frac{3 \minval}{2})}
{\exp(\frac{\minval \degmax}{2})},
\end{eqnarray*}
as claimed.
\hfill \end{proof}

Using Lemma~\ref{LemKeyDeg} and applying the bound~\eqref{EqnFanoOne}
from Lemma~\ref{LemFano} with $\NumModel = \frac{\pdim \degmax}{4}$
yields that for probability of error below $1/2$ and $\minval \degmax
\geq 2$, we require at least
\begin{eqnarray*}
\numobs & > & \frac{\log (\frac{\pdim \degmax}{4}
  -1)}{\symkull{\speparam{\graph^{st}}}{\speparam{\graph^{uv}}}} \;
\geq \; \frac{\exp(\frac{\degmax \minval}{2}) \; \log (\frac{\pdim
    \degmax}{4} -1)}{8 \degmax \minval \exp(\frac{3 \minval}{2})}
\end{eqnarray*}
samples.  Since $\exp(t/4) \geq t^2/16$ for all $t \geq 1$, we
certainly need at least $\numobs > \frac{\exp(\degmax \minval/4)
\degmax \minval \log (\frac{\pdim \degmax}{4} -1)}{128
\exp(\frac{3\minval}{2})}$ samples.  Since $\maxneigh = \degmax
\minval$ in this construction, we conclude that
\begin{eqnarray*}
\numobs & > & \frac{\exp(\maxneigh/4) \degmax \minval \log
(\frac{\pdim \degmax}{4} -1)}{128 \exp(\frac{3 \minval}{2})}
\end{eqnarray*}
samples are required, as claimed in Theorem~\ref{ThmNecDeg}. \\

\noindent {\bf{Ensemble C:}} Finally, we prove the third component in
the bound~\eqref{EqnNecDeg}.   In this case, we consider the ensemble
consisting of all graphs in $\Gcldeg$.  From Lemma~\ref{LemCard}(b),
we have
\begin{eqnarray*}
\log |\Gcldeg| & \geq & \frac{\degmax (\degmax+1)}{2} \, \log
\floor{\frac{\pdim}{\degmax+1}}!  \\
& \geq & \frac{\degmax (\degmax+1)}{2} \,
{\floor{\frac{\pdim}{\degmax+1}}} \; \log
\frac{\floor{\frac{\pdim}{\degmax+1}}}{e} \\
& \geq &  \frac{\degmax \pdim}{4} \log \frac{\pdim}{8 \degmax}.
\end{eqnarray*}
For this ensemble, it suffices to use a trivial upper bound on the
mutual information~\eqref{EqnMutInfo}, namely
\begin{eqnarray*}
\label{EqnTrivial}
I(\MyData; \graph) & \leq & H(\MyData) \; \leq \; \numobs \pdim,
\end{eqnarray*}
where the second bound follows since $\MyData$ is a collection of
$\numobs \pdim$ binary variables, each with entropy at most $1$.
Therefore, from the Fano bound~\eqref{EqnFano}, we conclude that the
error probability stays above $1/2$ if the sample size $\numobs$ is
upper bounded as $\numobs < \frac{\degmax}{8} \log \frac{\pdim}{8
\degmax}$, as claimed.

\subsection{Proof of Theorem~\ref{ThmNecEdge}}

We now turn to the proof of necessary conditions for the graph family
$\Gcledge$ with at most $\kdim$ edges.  As with the proof of
Theorem~\ref{ThmNecEdge}, it is based on applying Fano's inequality to
three ensembles of Markov random fields contained in
$\Gcledge(\minval, \maxneigh)$. \\

\noindent {\bf{Ensemble A:}} Note that the ensemble (A) previously
constructed in the proof of Theorem~\ref{ThmNecDeg} is also valid for
the family $\Gcledge(\minval, \maxneigh)$, and hence the
bound~\eqref{EqnBoundOne} is also valid for this family. \\\

\noindent {\bf{Ensemble B:}} For this ensemble, we choose the largest
integer $\mint$ such that $\kdim+1 \geq {\mint \choose 2}$.  Note that
we certainly have 
\begin{equation*}
{\mint \choose 2} \geq \lfloor \sqrt{\kdim}
\rfloor \; \geq \;  \frac{\sqrt{\kdim}}{2}.
\end{equation*}
  We then form a family of ${\mint \choose 2}$ graphs as
follows: (a) first form the complete graph $K_\mint$ on a subset of
$\mint$ vertices, and (b) for each $(s,t) \in K_\mint$, form the graph
$\graph^{st}$ by removing edge $(s,t)$ from $K_\mint$.  We form Markov
random fields on these graphs by setting $[\speparam{\graph^{st}}]_{wz}
= \minval$ if $(w,z) \in \edge(\graph^{st})$, and setting it to zero
otherwise.

\blems
\label{LemSymkullKreg}
For all distinct model
pairs $\speparam{\graph^{st}}$ and $\speparam{\graph^{uv}}$, we have
\begin{eqnarray}
\label{EqnSymkullKreg}
\symkull{\speparam{\graph^{st}}}{\speparam{\graph^{uv}}} & \leq &
\frac{16 \maxneigh \exp(\frac{5 \minval}{2}) \;
\sinh(\minval)}{\exp(\frac{\maxneigh}{2})}.
\end{eqnarray}
\elems
\begin{proof}
We begin by claiming for any pair $(s,t) \neq (u,v)$, the distribution
$\mprob_{\speparam{\graph^{uv}}}$ (i.e., corresponding to the subgraph
that does \emph{not} contain edge $(u,v)$) satisfies
\begin{eqnarray}
\label{EqnAlon}
\frac{\mprob_{\speparam{\graph^{uv}}}[X_s X_t =
+1]}{\mprob_{\speparam{\graph^{uv}}}[X_s X_t = -1]} & \leq & \exp(2
\minval) \; \frac{\mprob_{\speparam{\graph^{st}}}[X_s X_t =
+1]}{\mprob_{\speparam{\graph^{st}}}[X_s X_t = -1]} \; = \; \exp(2
\minval) \frac{\myq}{1-\myq},
\end{eqnarray}
where we have re-introduced the convenient shorthand $\myq =
\mprob_{\speparam{\graph^{st}}}[X_s X_t = 1]$ from
Lemma~\ref{LemKeySep}.

To prove this claim, let $\mprob_{\eparam}$ be the distribution that
contains all edges in the complete subgraph $K_\mint$, each with
weight $\minval$.  Let $Z(\eparam)$ and $Z(\speparam{\graph^{uv}})$ be
the normalization constants associated with $\mprob_\eparam$ and
$\mprob_{\speparam{\graph^{uv}}}$ respectively.  Now since $\minval >
0$ by assumption, by the FKG inequality~\cite{Alon}, we have
\begin{eqnarray*}
\frac{\mprob_{\speparam{\graph^{uv}}}[X_s X_t =
+1]}{\mprob_{\speparam{\graph^{uv}}}[X_s X_t = -1]} & \leq &
\frac{\mprob_{\eparam}[X_s X_t = +1]}{\mprob_\eparam[X_s X_t = -1]}
\end{eqnarray*}
We now apply the definition of $\mprob_\eparam$ and expand the
right-hand side of this expression, recalling the fact that the model
$\mprob_{\speparam{\graph^{st}}}$ does \emph{not} contain the edge
$(s,t)$.  Thus we obtain
\begin{eqnarray*}
\frac{\mprob_{\eparam}[X_s X_t = +1]}{\mprob_\eparam[X_s X_t = -1]} &
= & \frac{\exp(\minval) \frac{Z(\speparam{\graph^{st}})}{Z(\eparam)}  \; \:
\mprob_{\speparam{\graph^{st}}}[X_s X_t = +1]}
{\exp(-\minval) \frac{Z(\speparam{\graph^{st}})}{Z(\eparam)}  \; \:
\mprob_{\speparam{\graph^{st}}}[X_s X_t = -1]}  \\
& = & \exp(2 \minval) \; \frac{ \mprob_{\speparam{\graph^{st}}}[X_s
X_t = +1]} { \mprob_{\speparam{\graph^{st}}}[X_s X_t = -1]},
\end{eqnarray*}
which establishes the claim~\eqref{EqnAlon}.

Finally, from the representation~\eqref{EqnNewSymKull} for the symmetrized
Kullback-Leibler divergence and the definition of the models,
\begin{eqnarray*}
\symkull{\speparam{\graph^{st}}}{\speparam{\graph^{uv}}} & = & \minval
\big \{ \Exs_{\speparam{\graph^{st}}}[X_u X_v] -
\Exs_{\speparam{\graph^{uv}}}[X_u X_v] \big \} + \minval \big \{
\Exs_{\speparam{\graph^{uv}}}[X_s X_t] -
\Exs_{\speparam{\graph^{st}}}[X_s X_t] \big \} \\
& = & 2 \minval \big \{ \Exs_{\speparam{\graph^{uv}}}[X_s X_t] -
\Exs_{\speparam{\graph^{st}}}[X_s X_t] \big \},
\end{eqnarray*}
where we have used the symmetry of the two terms.  Continuing on, we
observe the decomposition \mbox{$\Exs_{\speparam{\graph^{st}}}[X_s
X_t] = 1-2 \mprob_{\speparam{\graph^{st}}}[X_s X_t = -1]$,} and using
the analogous decomposition for the other expectation, we obtain
\begin{eqnarray*}
\symkull{\speparam{\graph^{st}}}{\speparam{\graph^{uv}}} & = & 4
\minval \big \{\mprob_{\speparam{\graph^{st}}}[X_s X_t = -1]
-\mprob_{\speparam{\graph^{uv}}}[X_s X_t = -1] \big \} \\
& = & 4 \minval \Big \{ \frac{1}{
\frac{\mprob_{\speparam{\graph^{st}}}[X_s X_t = +1]}
{\mprob_{\speparam{\graph^{st}}}[X_s X_t = -1]} +1 } - \frac{1}{
\frac{\mprob_{\speparam{\graph^{uv}}}[X_s X_t = +1]}
{\mprob_{\speparam{\graph^{uv}}}[X_s X_t = -1]} +1 } \Big \} \\
& \stackrel{(a)}{\leq} & 4 \minval \Big \{
\frac{1}{\frac{\myq}{1-\myq} +1} - \frac{1}{ \exp(2 \minval)
\frac{\myq}{1-\myq} +1} \Big \} \\
& = & \frac{4 \minval}{\frac{\myq}{1-\myq}} \Big \{ \frac{1}{1 +
\frac{1-\myq}{\myq}} - \frac{1}{\exp(2 \minval) + \frac{1-\myq}{\myq}}
\Big \} \\
& = & \frac{4 \minval \, (\exp(2\minval) - 1)}{\frac{\myq}{1-\myq}} \;
\; \frac{1}{\big[1 +\frac{1-\myq}{\myq}\big] \; \big[\exp(2 \minval) +
\frac{1-\myq}{\myq} \big]}
\end{eqnarray*}
where in obtaining the inequality (a), we have applied the
bound~\eqref{EqnAlon} and recalled our shorthand notation \mbox{$\myq
= \mprob_{\speparam{\graph^{st}}}[X_s X_t = +1]$.}  Since $\minval >
0$ and $(1-\myq)/\myq \geq 0$, both terms in the denominator of the
second term are at least one, so that we conclude that
$\symkull{\speparam{\graph^{st}}}{\speparam{\graph^{uv}}} \leq \frac{4
\minval \, (\exp(2\minval) - 1)}{\frac{\myq}{1-\myq}}$.

Finally, applying the lower bound~\eqref{EqnInter} from
Lemma~\ref{LemKeySep} on the ratio $\myq/(1-\myq)$, we obtain that
\begin{eqnarray*}
\symkull{\speparam{\graph^{st}}}{\speparam{\graph^{uv}}} & \leq &
\frac{4 \minval \, (\exp(2\minval) -1) \;
(\mint+1)}{\exp(\frac{\maxneigh}{2} - \frac{3}{2} \minval)} \\
& \leq & \frac{16 \maxneigh \exp(\frac{5 \minval}{2}) \;
\sinh(\minval)}{\exp(\frac{\maxneigh}{2})},
\end{eqnarray*}
where we have used the fact that $\minval \, (\mint+1) \leq 2 \mint
\minval \; = \; 2 \maxneigh$.  
\hfill \end{proof}

By combining Lemma~\ref{LemSymkullKreg} with Lemma~\ref{LemFano}(b),
we conclude that for correctness with probability at least $1/2$, the
sample size $\numobs$ must be at least
\begin{eqnarray*}
\numobs & > & \frac{ \exp(\frac{\maxneigh}{2}) \; \log \frac{{\mint
\choose 2}}{4}}{32 \maxneigh \exp(\frac{5 \minval}{2}) \;
\sinh(\minval)} \; \geq \; \frac{ \exp(\frac{\maxneigh}{2}) \; \log
(\kdim/8)}{64 \maxneigh \exp(\frac{5 \minval}{2}) \; \sinh(\minval)},
\end{eqnarray*}
as claimed in Theorem~\ref{ThmNecEdge}.

\section{Proofs of sufficient conditions}
\label{SecSufProof}

We now turn to the proofs of the sufficient conditions given in
Theorems~\ref{ThmSufDeg} and~\ref{ThmSufEdge}, respectively, for the
classes $\Gcldeg$ and $\Gcledge$.  In both cases, our method involves a
direct analysis of a maximum likelihood (ML) decoder, which searches
exhaustively over all graphs in the given class, and computes the
model with highest likelihood.  We begin by describing this ML decoder
and providing a standard large deviations bound that governs its
performance.  The remainder of the proof involves more delicate
analysis to lower bound the error exponent in the large deviations
bound in terms of the minimum edge weight $\minval$ and other
structural properties of the distributions.

\subsection{ML decoding and large deviations bound}

Given a collection $\MyData = \{\Xsam{1}, \ldots, \Xsam{\numobs}\}$
of $\numobs$ i.i.d. samples,  its (rescaled) likelihood with respect to model
$\mprob_\eparam$ is given by
\begin{eqnarray}
\like_\eparam(\MyData) & \defn & \frac{1}{\numobs} \sum_{i=1}^n
\log \mprob_\eparam[\Xsam{i}].
\end{eqnarray}
For a given graph class $\Gclass$ and an associated set of graphical
models $\{ \speparam{\graph} \; \mid \; \graph \in \Gclass \}$, the
\emph{maximum likelihood decoder} is the mapping $\likedecode:
\StateSp \rightarrow \Gclass$ defined by 
\begin{eqnarray}
\label{EqnML}
\likedecode(\MyData) & = & \arg \max_{\graph \in \Gclass}
\like_{\speparam{\graph}}(\MyData).
\end{eqnarray}
(If the maximum is not uniquely achieved, we choose some graph
$\graph$ from the set of models that attains the maximum.)

Suppose that the data is drawn from model $\mprob_{\speparam{\graph}}$
for some $\graph \in \Gclass$.  Then the ML decoder $\likedecode$
fails only if there exists some other $\speparam{\graph'} \neq
\speparam{\graph}$ such that $\like_{\speparam{\graph'}}(\MyData) \geq
\like_{\speparam{\graph}}(\MyData)$.  (Note that we are being
conservative by declaring failure when equality holds).  Consequently,
by union bound, we have
\begin{eqnarray*}
\mprob_{\eparam}[\likedecode(\MyData) \neq \graph] & \leq &
\sum_{\graph' \in \Gclass \backslash \graph} \mprob \big[
\ell_{\speparam{\graph'}}(\MyData) \geq
\ell_{\speparam{\graph}}(\MyData) \big]
\end{eqnarray*}
Therefore, in order to provide sufficient conditions for the error
probability of the ML decoder to vanish, we need to provide
an appropriate large deviations bound.

\begin{lemma} 
\label{LemLargeDev}
 Given $\numobs$ i.i.d. samples $\MyData = \{\Xsam{1}, \ldots,
\Xsam{\numobs} \}$ from $\mprob_{\speparam{\graph}}$, for any
$\graph' \neq \graph$, we have
\begin{eqnarray}
\label{EqnLargeDev}
\mprob \big[ \ell_{\speparam{\graph'}}(\MyData) \geq
\ell_{\speparam{\graph}}(\MyData) \big] & \leq & \exp \big(-
\frac{\numobs}{2} \, \weirdkull{\speparam{\graph}}{\speparam{\graph'}},
\end{eqnarray}
where the distance $S$ was defined previously~\eqref{EqnWeirdKull}.
\end{lemma}
\begin{proof}
So as to lighten notation, let us write $\eparam = \speparam{\graph}$
and $\eparam' = \speparam{\graph'}$.  We apply the Chernoff bound to
the random variable $V = \ell_{\eparam'}(\MyData) -
\ell_{\eparam}(\MyData)$, thereby obtaining that
\begin{eqnarray*}
\frac{1}{\numobs} \log \mprob_{\eparam}[V \geq 0] & \leq &
\frac{1}{\numobs} \inf_{s>0} \log \Exs_{\eparam} \big[ \exp(s V) \big]
\\
& = & \inf_{s > 0} \sum_{x \in \{-1,+1\}^\pdim}
\big[\mprob_\eparam(x)\big]^{1-s} \; \big[\mprob_{\eparam'}(x)\big]^s \\
& \leq & \log Z\big( \eparam/2 + \eparam'/2) - \frac{1}{2} \log
Z(\eparam) - \frac{1}{2} \log Z(\eparam'),
\end{eqnarray*}
where $Z(\eparam)$ denotes the normalization constant associated with
the Markov random field $\mprob_\eparam$, as defined in
equation~\eqref{EqnDefnZ}.  The claim then follows by applying the
representation~\eqref{EqnNewWeirdKull} of
$\weirdkull{\eparam}{\eparam'}$.

\hfill \end{proof}
%


\subsection{Lower bounds based on matching}

In order to exploit the large deviations claim in
Lemma~\ref{LemLargeDev}, we need to derive lower bounds on the
divergence $\weirdkull{\speparam{\graph}}{\speparam{\graph'}}$ between
different models.  Intuitively, it is clear that this divergence is
related to the discrepancy of the edge sets of the two graph.  The
following lemma makes this intuition precise.  We first recall some
standard graph-theoretic terminology: a \emph{matching} of a graph
$\graph = (\vertex, \edge)$ is a subgraph $H$ such that each vertex in
$H$ has degree one.  The \emph{matching number} of $\graph$ is the
maximum number of edges in any matching of $\graph$.

\begin{lemma}
\label{LemMatch}
Given two distinct graphs $\graph = (\vertex, \edge)$ and $\graph' =
(\vertex, \edge')$, let $\matchnum(\graph, \graph')$ be the matching
number of the graph with edge set
\begin{eqnarray*}
\edge \Delta \edge' & \defn & (\edge \backslash \edge') \, \cup \,
(\edge' \backslash \edge').
\end{eqnarray*}
Then for any pair of parameter vectors $\speparam{\graph}$ and
$\speparam{\graph'}$ in $\Gclass(\minval, \maxneigh)$, we have
\begin{eqnarray}
\label{EqnMatchLow}
\weirdkull{\speparam{\graph}}{\speparam{\graph'}} & \geq &
\frac{\matchnum(\graph, \graph')}{3 \, \exp(2 \maxneigh)+1} \;
\sinh^{2}(\frac{\minval}{4}).
\end{eqnarray}
\end{lemma}
\begin{proof}
Some comments on notation before proceeding: we again adopt the
shorthand notation $\eparam = \speparam{\graph}$ and $\eparam' =
\speparam{\graph'}$.  In this proof, we use $e_j$ to denote either a
particular edge, or the set of two vertices that specify the edge,
depending on the context.  Given any subset $\Aset \subset \vertex$,
we use $x_\Aset = \{x_s, \; s \in \Aset \}$ to denote the collection
of variables indexed by $\Aset$.

Given any edge $e = (u,v)$ with $u \notin \Aset$ and $v \notin \Aset$,
we define the conditional distribution
\begin{eqnarray}
\label{EqnCond}
\mprob^{e}_{\eparam[x_\Aset]}(x_u, x_v) & = &
\frac{\mprob_{\eparam}(x_u, x_v, x_\Aset)}{\mprob_{\eparam}(x_\Aset)}
\end{eqnarray}
over the random variables $x_e = (x_u, x_v)$ indexed by the edge.
Finally, we use
\begin{eqnarray}
\label{EqnDefnSubWeird}
\newweird{e}{x_\Aset}{\eparam}{\eparam'} & \defn &
\bigkull{\mprob^{e}_{\frac{\eparam[x_\Aset] + \eparam'[x_\Aset]}{2}}}
{\mprob^{e}_{\eparam[x_\Aset]}} +
\bigkull{\mprob^{e}_{\frac{\eparam[x_\Aset] + \eparam'[x_\Aset]}{2}}}
{\mprob^{e}_{\eparam'[x_\Aset]}}
\end{eqnarray}
to denote the divergence~\eqref{EqnWeirdKull} applied to the conditional
distributions of $(X_u, X_v \, \mid X_\Aset = x_\Aset)$.

With this notation, let $M \subset \edge \Delta \edge'$ be the subset
of edges in some \emph{maximal} matching of the graph with edge set
$\edge \Delta \edge'$; concretely, let us write $M = \{e_1, \ldots,
e_m\}$, and denote by $\vertex \backslash M$ the subset of vertices
that are \emph{not} involved in the matching.  Note that since
$\plweird$ is a combination of Kullback-Leibler (KL) divergences, the
usual chain rule for KL divergences~\cite{Cover} also applies to it.
Consequently, we have
\begin{eqnarray*}
\weirdkull{\eparam}{\eparam'} & \geq & \sum_{\ell=1}^{\matchnum} \quad
\sum_{x_{\vertex \backslash M}, x_{e_1}, \ldots x_{e_{\ell-1}} }
\mprob_{\frac{\eparam + \eparam'}{2}} (x_{\vertex \backslash M},
x_{e_1}, \ldots, x_{e_{\ell-1}}) \; \newweird{e_\ell}{x_{S_{\ell-1}}}
{\eparam}{\eparam'},
\end{eqnarray*}
where for each $\ell$, we are conditioning on the set of variables
$x_{S_{\ell-1}} \defn \big(x_{\vertex \backslash M}, x_{e_1}, \ldots,
x_{e_{\ell-1}} \big)$.  Finally, from Lemma~\ref{LemKLTech} in
Appendix~\ref{AppLemKLTech}, for all $\ell = 1, \ldots, \matchnum$ and
all values of $x_{S_{\ell-1}}$, we have
\begin{eqnarray*}
\newweird{e_\ell}{x_{S_{\ell-1}}} {\eparam}{\eparam'} & \geq &
\frac{1}{3 \exp(2 \maxneigh) + 1} \sinh^2(\frac{\minval}{4}),
\end{eqnarray*}
from which the claim follows.

\hfill \end{proof}


\subsection{Proof of Theorem~\ref{ThmSufDeg}(a)}

We first consider distributions belonging to the class
$\Gcldeg(\minval, \maxneigh)$, where $\minval$ is the minimum absolute
value of any non-zero edge weight, and $\maxneigh$ is the maximum
neighborhood weight~\eqref{EqnDefnMaxNeigh}.  Consider a pair of
graphs $\graph$ and $\graph'$ in the class $\Gcldeg$ that differ in
$\ell = |\edge \, \Delta \edge'|$ edges.  Since both graphs have
maximum degree at most $\degmax$, we necessarily have a matching
number $\matchnum(\graph, \graph') \geq \frac{\ell}{4 \, \degmax}$.
Note that the parameter $\ell = |\edge \, \Delta \edge'|$ can range
from $1$ all the way up to $\degmax \pdim$, since a graph with maximum
degree $\degmax$ has at most $\frac{\degmax \pdim}{2}$ edges.

Now consider some fixed graph $\graph \in \Gcldeg$ and associated
distribution $\mprob_{\speparam{\graph}} \in \Gcldeg$; we upper bound
the error probability $\mprob_{\speparam{\graph}}[\decopt(\MyData)
\neq \graph]$.  For each $\ell = 1, 2, \ldots, \degmax \pdim$, there
are at most ${{\pdim \choose 2} \choose \ell}$ models in $\Gcldeg$
with mismatch $\ell$ from $\graph$.  Therefore, applying the union bound,
the large deviations bound in Lemma~\ref{LemLargeDev}, and the lower
bound in terms of matching from Lemma~\ref{LemMatch}, we obtain
\begin{eqnarray*}
\mprob_{\speparam{\graph}}[\decopt(\MyData) \neq \graph] & \leq &
\sum_{\ell=1}^{\pdim \degmax} {{\pdim \choose 2} \choose \ell} \; \exp
\Big \{ -\numobs \; \frac{\ell/(4 \degmax) }{3 \exp(2 \maxneigh)+1} \,
\sinh^{2}(\frac{\minval}{2}) \Big \}\\
& \leq & \pdim \degmax \; \max_{\ell = 1, \ldots, \pdim \degmax } \exp
 \Big \{ \log {{\pdim \choose 2} \choose \ell} -\numobs \;
 \frac{\ell/(4 \degmax) }{3 \exp(2 \maxneigh)+1} \,
 \sinh^{2}(\frac{\minval}{2}) \Big \} \\
& \leq & \max_{\ell = 1, \ldots, \pdim \degmax} \, \exp \Big \{ \log (\pdim \degmax) + \ell \log \pdim 
 - \numobs \frac{\ell/(4 \degmax) }{3 \exp(2 \maxneigh)+1} \,
 \sinh^{2}(\frac{\minval}{2}) \Big \}.
\end{eqnarray*}
This probability is at most $\delta$ under the given conditions on
$\numobs$ in the statement of Theorem~\ref{ThmSufDeg}(a).

\subsection{Proof of Theorem~\ref{ThmSufEdge}}

Next we consider the class $\Gcledge$ of graphs with at most $\kdim$
edges. Given some fixed graph $\graph \in \Gcledge$, consider some
other graph $\graph' \in \Gcledge$ such that the set $\edge \, \Delta
\, \edge'$ has cardinality $\matchnum$.  We claim that for each
$\matchnum = 1, 2, \ldots, 2 \kdim$, the number of such graphs
is at most 

To verify this claim, recall the notion of a \emph{vertex cover} of a
set of edges, namely a subset of vertices such that each edge in the
set is incident to at least one vertex of the set.  Note also that the
vertices involved in any maximal matching form a vertex cover.
Consequently, any maximal matching over the edge set $\edge \, \Delta
\, \edge'$ of cardinality $\matchnum$ be described in the following
(suboptimal) fashion: (i) first specify which of the $\kdim$ edges in
$\edge$ are missing in $\edge'$; (ii) describe which of the at most $2
\matchnum$ vertices belong to the vertex cover defined by the maximal
matching; and (iii) describe the subset of at most $\kdim$ vertices
that are connected to it.  This procedure yields at most
\mbox{$2^{\kdim} \pdim^{2 \matchnum} \pdim^{2 \, \kdim \, \matchnum} =
2^\kdim \pdim^{2 \matchnum \, (\kdim +1)}$} possibilities, as claimed.

Consequently, applying the union bound, the large deviations bound in
Lemma~\ref{LemLargeDev}, and the lower bound in terms of matching from
Lemma~\ref{LemMatch}, we obtain
\begin{eqnarray*}
\mprob_{\speparam{\graph}}[\decopt(\MyData) \neq \graph] & \leq &
\sum_{\matchnum=1}^{\kdim} 2^{\kdim} \pdim^{2 \, \matchnum \,
(\kdim+1)} \exp \Big \{ - \numobs \frac{\matchnum}{3 \, \exp(2
\maxneigh)+1} \; \sinh^{2}(\frac{\minval}{2}) \Big \} \\
 & \leq & \kdim \max_{\matchnum = 1, \ldots, \kdim } \; \exp \Big \{
\kdim + 2 \matchnum (\kdim + 1) \log \pdim 
- \numobs \frac{\matchnum}{3 \, \exp(2
\maxneigh)+1} \; \sinh^{2}(\frac{\minval}{2}) \Big \}.
\end{eqnarray*}
This probability is less than $\delta$ under the conditions of
Theorem~\ref{ThmSufEdge}, which completes the proof.

\subsection{Proof of Theorems~\ref{ThmSufDeg}(b) and~\ref{ThmSufEdge}(b)}

Finally, we prove the sufficient conditions given in
Theorem~\ref{ThmSufDeg}(b) and~\ref{ThmSufEdge}(b), which do not
assume that the decoder knows the parameter vector $\speparam{\graph}$
for each graph $\graph \in \Gcldeg$.  In this case, the simple ML
decoder~\eqref{EqnML} cannot be applied, since it assumes knowledge of
the model parameters $\speparam{\graph}$ for each graph $\graph \in
\Gcldeg$.  A natural alternative would be the generalized likelihood
ratio approach, which would maximize the likelihood over each model
class, and then compare the maximized likelihoods. Our proof of
Theorem~\ref{ThmSufDeg}(b) is based on minimizing the distance
between the empirical and model mean parameters in the $\ell_\infty$
norm, which is easier to analyze.

\subsubsection{Decoding from mean parameters}

We begin by describing the graph decoder used to establish the
sufficient conditions of Theorem~\ref{ThmSufDeg}(b).  For any
parameter vector $\eparam \in \real^{{\pdim \choose 2}}$, let
$\meanpar(\eparam) \in \real^{{\pdim \choose 2}}$ represent the
associated set of mean parameters, with element $(s,t)$ given by
$[\meanpar(\eparam)]_{st} \defn \Exs_{\eparam}[X_s X_t]$.  Given a
data set $\MyData = \{ \Xsam{1}, \ldots, \Xsam{\numobs}\}$, the
\emph{empirical mean parameters} are given by
\begin{eqnarray}
\label{EqnDefnEmpMean}
\estim{\meanpar}_{st} & \defn & \frac{1}{\numobs} \sum_{i=1}^\numobs
\Xsam{i}_s \Xsam{i}_t.
\end{eqnarray}

For a given graph $\graph = (\vertex, \edge)$, let $\Theta_{\minval,
\maxneigh}(\graph) \subset \real^{{\pdim \choose 2}}$ be a subset of
exponential parameters that respect the graph structure---viz.
\begin{enumerate}
\item[(a)]  we have $\eparam_{uv} = 0$ for all $(u,v) \notin \edge$;
\item[(b)] for all edges $(s,t) \in \edge$, we have $|\eparam_{st}|
\geq \minval$, and
\item[(c)] for all vertices $s \in \vertex$, we have $\sum_{t \in
\Neigh(s)} |\eparam_{st}| \leq \maxneigh$.
\end{enumerate}
For any graph $\graph$ and set of mean parameters $\meanpar \in
\real^{{\pdim \choose 2}}$, we define a projection-type distance via
$\Jcost_\graph(\meanpar) = \min_{\eparam \in \Theta_{\minval,
\maxneigh}(\graph)} \| \meanpar - \meanpar(\eparam) \|_\infty$.

We now have the necessary ingredients to define a graph decoder
$\dectwo: \StateSp^\numobs \rightarrow \Gcldeg$; in particular, it is
given by
\begin{eqnarray}
\label{EqnDefnDectwo}
\dectwo(\MyData) & \defn & \arg \min_{\graph \in \Gcldeg}
\Jcost_\graph(\estim{\meanpar}),
\end{eqnarray}
where $\estim{\meanpar}$ are the empirical mean parameters previously
defined~\eqref{EqnDefnEmpMean}.  (If the minimum~\eqref{EqnDefnDectwo}
is not uniquely achieved, then we choose some graph that achieves the
minimum.)

\newcommand{\Delg}[2]{\ensuremath{\Delta(#1 \, ; \, #2)}}

\subsubsection{Analysis of decoder}

Suppose that the data are sampled from $\mprob_{\speparam{\graph}}$
for some fixed but known graph $\graph \in \Gcldeg$, and parameter
vector $\speparam{\graph} \in \Theta_{\minval, \maxneigh}(\graph)$.
Note that the graph decoder $\dectwo$ can fail only if there exists
some other graph $\graph'$ such that the difference
$\Delg{\graph'}{\graph} \defn \Jcost_{\graph'}(\estim{\meanpar}) -
\Jcost_\graph(\estim{\meanpar})$ is not positive.  (Again, we are
conservative in declaring failure if there are ties.)

Let $\eparam'$ denote some element of $\Theta_{\minval,
\maxneigh}(\graph')$ that achieves the minimum defining
$\Jcost_{\graph'}(\estim{\meanpar})$, so that
$\Jcost_{\graph'}(\estim{\meanpar}) = \| \estim{\meanpar} -
\meanpar(\eparam') \|_\infty$.  Note that by the definition of
$\Jcost_\graph$, we have $\Jcost_\graph(\estim{\graph}) \leq \|
\estim{\meanpar} - \speparam{\graph} \|_\infty$, where $\speparam{\graph}$
are the parameters of the true model.  Therefore, by the definition
of $\Del{\graph'}{\graph}$, we have
\begin{eqnarray}
\Delg{\graph'}{\graph} & \geq & \| \estim{\meanpar} -
\meanpar(\eparam') \|_\infty - \| \estim{\meanpar} -
\meanpar(\speparam{\graph}) \|_\infty \notag \\
\label{EqnLowerBound}
& \geq & \| \meanpar(\eparam') - \meanpar(\speparam{\graph}) \|_\infty
- 2 \| \estim{\meanpar} - \meanpar(\speparam{\graph}) \|_\infty,
\end{eqnarray}
where the second inequality applies the triangle inequality.

Therefore, in order to prove that $\Delg{\graph'}{\graph}$ is
positive, it suffices to obtain an upper bound on $\| \estim{\meanpar}
- \meanpar(\speparam{\graph}) \|_\infty$, and a lower bound on
$\|\meanpar(\eparam') - \meanpar(\speparam{\graph})\|_\infty$, where
$\eparam'$ ranges over $\Theta_{\minval, \maxneigh}(\graph')$.  With
this perspective, let us state two key lemmas.  We begin with the
deviation between the sample and population mean parameters:
\begin{lemma}[Elementwise deviation]  
\label{LemElementwise}
Given $\numobs$ i.i.d. samples drawn from
$\mprob_{\speparam{\graph}}$, the sample mean parameters
$\estim{\meanpar}$ and population mean parameters
$\meanpar(\speparam{\graph})$ satisfy the tail bound
\begin{eqnarray*}
\mprob[\| \estim{\meanpar} - \meanpar(\speparam{\graph}) \|_\infty
 \geq t] & \leq & 2 \exp \big(- \numobs \frac{t^2}{2} + 2 \log \pdim
 \big).
\end{eqnarray*}
This probability is less than $\delta$ for $t \geq \sqrt{\frac{4 \log
\pdim + \log(2/\delta)}{\numobs}}$.
\end{lemma}
Our second lemma concerns the separation of the mean parameters of
models with different graph structure:

\begin{lemma}[Pairwise separations]
\label{LemMeanSep}
Consider any two graphs $\graph = (\vertex, \edge)$ and $\graph' =
(\vertex, \edge')$, and an associated set of model parameters
$\speparam{\graph} \in \Theta_{\minval, \maxneigh}(\graph)$ and
$\speparam{\graph'} \in \Theta_{\minval, \maxneigh}(\graph')$.  Then
for all edges $(s,t) \in \edge\backslash \edge' \cup \edge' \backslash
\edge$,
\begin{eqnarray*}
\max_{\substack{u \in \{s,t \}, v \in \vertex} } \big |
\Exs_{\speparam{\graph}} [X_u X_v] - \Exs_{\speparam{\graph'}} [X_u
X_v] \big| & \geq & \frac{\sinh^{2}(\minval/4)}{2 \maxneigh \, \big(3
\exp(2 \maxneigh) +1 \big)}.
\end{eqnarray*}
\end{lemma} 
\noindent We provide the proofs of these two lemmas in
Sections~\ref{AppLemElementwise} and~\ref{AppLemMeanSep} below.

Given these two lemmas, we can complete the proofs of
Theorem~\ref{ThmSufDeg}(b) and Theorem~\ref{ThmSufEdge}(b).  Using the
lower bound~\eqref{EqnLowerBound}, with probability greater than
$1-\delta$, we have
\begin{eqnarray*}
\Delg{\graph'}{\graph} & \geq & 
\frac{\sinh^{2}(\minval/4)}{\maxneigh
\, \big(3 \exp(2 \maxneigh+1 \big)}
 - 2 \sqrt{\frac{4 \log \pdim +
\log(2/\delta)}{\numobs}}.
\end{eqnarray*}
This quantity is positive as long as 
\begin{eqnarray*}
\numobs & > & \Big[\frac{\maxneigh \, \big(3 \exp(2 \maxneigh+1
\big)}{\sinh^{2}(\minval/4)}\Big]^2 \big \{ 16 \log \pdim + 4
\log(2/\delta) \big \},
\end{eqnarray*}
which completes the proof.

It remains to prove the auxiliary lemmas used in the proof.

\subsubsection{Proof of Lemma~\ref{LemElementwise}}
\label{AppLemElementwise}

This claim is an elementary consequence of the Hoeffding bound.  By
definition, for each pair $(s,t)$ of distinct vertices, we have
\begin{eqnarray*}
\estim{\meanpar}_{st} - [\meanpar(\speparam{\graph})]_{st} & = &
\frac{1}{\numobs} \sum_{i=1}^\numobs \Xsam{i}_s \Xsam{i}_t -
\Exs_{\speparam{\graph}}[X_s X_t],
\end{eqnarray*}
which is the deviation of a sample mean from its expectation.  Since
the random variables $\{ \Xsam{i}_s \Xsam{i}_t \}_{i=1}^n$ are
i.i.d. and lie in the interval $[-1, +1]$, an application of
Hoeffding's inequality~\cite{Hoeffding63} yields that
\begin{eqnarray*}
\mprob[\big| \estim{\meanpar}_{st} -
[\meanpar(\speparam{\graph})]_{st} \big | \geq t] & \leq & 2 \, \exp
\big( - \numobs t^2/2).
\end{eqnarray*}
The lemma follows by applying union bound over all ${\pdim \choose 2}$
edges of the graph, and the fact that $\log {\pdim \choose 2} \leq 2
\log \pdim$.

\subsubsection{Proof of Lemma~\ref{LemMeanSep}}
\label{AppLemMeanSep}

\newcommand{\Spec}{\ensuremath{C}}
\newcommand{\Qprob}{\ensuremath{\mathbb{Q}}}
\newcommand{\Yset}{\ensuremath{\mathcal{Y}}}
\newcommand{\newSet}{\ensuremath{\Yset^*(x_U)}}

The proof of this claim is more involved.  Let $(s,t)$ be an edge in
$\edge \backslash \edge'$, and let $\Spec$ be the set of all other
vertices that are adjacent to $s$ or $t$ in either graphs---namely,
the set
\begin{eqnarray*}
\Spec & \defn & \big \{ u \in \vertex \, \mid \, (u,s) \text{ or }
(u,t) \in \edge \cup \edge' \big \} \; = \; \big(\Neigh(s) \cup
\Neigh(t) \big) \backslash \{s, t\}.
\end{eqnarray*}
Our approach is to condition on the variables $x_\Spec = \{ x_u, u \in
\Spec\}$, and consider the two conditional distributions over the pair
$(X_s, X_t)$, defined by $\mprob_\eparam$ and $\mprob_{\eparam'}$
respectively.  In particular, for any subset $S \subset \vertex$, let
us define the unnormalized distribution
\begin{eqnarray}
\label{EqnQprob}
\Qprob_{\eparam}(x_S) & \defn & \sum_{x_a, \; a \notin S}
\exp \big(\sum_{(u,v) \in \edge} \eparam_{uv} x_u x_v \big),
\end{eqnarray}
obtained by summing out all variables $x_a$ for $a \notin S$.  With
this notation, we can write the conditional distribution of $(X_s,
X_t)$ given $\{X_\Spec = x_\Spec\}$ as
\begin{eqnarray}
\mprob_{\eparam[x_\Spec]}(x_s, x_t) & = & \frac{\Qprob_{\eparam}(x_s,
x_t, x_\Spec)}{\Qprob_\eparam(x_\Spec)}.
\end{eqnarray}
As reflected in our choice of notation, for each fixed $x_\Spec$, the
distribution~\eqref{EqnCond} can be viewed as a Ising model over the
pair $(X_s, X_t)$ with exponential parameter $\eparam[x_\Spec]$.  We
define the unnormalized distributions $\Qprob_{\eparam'[x_S]}$ and the
conditional distributions $\mprob_{\eparam'[x_\Spec]}$ in an analogous
manner.

Our approach now is to study the divergence
$\weirdkull{\eparam[x_\Spec]}{\eparam'[x_\Spec]}$ between the
conditional distributions induced by $\mprob_\eparam$ and
$\mprob_{\eparam'}$. Using Lemma~\ref{LemKLTech} from
Appendix~\ref{AppLemKLTech}, for each choice of $x_\Spec$, we have
\mbox{$\weirdkull{\eparam[x_\Spec]}{\eparam'[x_\Spec]} \geq
\frac{\sinh^2(\frac{\lambda}{4})}{3 \exp(2 \maxneigh) + 1}$,} and
hence
\begin{eqnarray}
\label{EqnAve}
\Exs_{\frac{\eparam + \eparam'}{2}}
\big[\weirdkull{\eparam[X_\Spec]}{\eparam'[X_\Spec]} \big] & \geq &
\frac{1}{3 \exp(2 \maxneigh) + 1} \sinh^2(\frac{\lambda}{4}),
\end{eqnarray}
where the expectation is taken under the model $\mprob_{\frac{\eparam
+ \eparam'}{2}}$.  Some calculation shows that
\begin{eqnarray*}
\Exs_{\frac{\eparam + \eparam'}{2}}
\big[\weirdkull{\eparam[X_\Spec]}{\eparam'[X_\Spec]} \big] & = &
\Exs_{\frac{\eparam + \eparam'}{2}} \biggr[ \log
\frac{\Qprob_\eparam(X_\Spec)
}{\Qprob{\frac{\eparam+\eparam'}{2}}(X_\Spec)} \biggr] +
\Exs_{\frac{\eparam + \eparam'}{2}} \biggr[ \log
\frac{\Qprob_{\eparam'}(X_\Spec)
}{\Qprob{\frac{\eparam+\eparam'}{2}}(X_\Spec)} \biggr].
\end{eqnarray*}
Applying Jensen's inequality yields that
\begin{eqnarray*}
\Exs_{\frac{\eparam + \eparam'}{2}} \biggr[ \log
\frac{\Qprob_\eparam(X_\Spec)
}{\Qprob{\frac{\eparam+\eparam'}{2}}(X_\Spec)} \biggr] & \leq & \log
\biggr[ \frac{1}{\sum_{x_\Spec}
\Qprob_{\frac{\eparam+\eparam'}{2}}(x_\Spec)} \; \sum_{x_\Spec}
\Qprob_{\frac{\eparam+\eparam'}{2}}(x_\Spec)
\frac{\Qprob_{\eparam}(x_\Spec)}{\Qprob_{\frac{\eparam'+
\eparam}{2}}[x_\Spec]} \biggr] \\
& = & \log \frac{\sum_{x_\Spec} \Qprob_{\eparam}(x_\Spec)}
{\sum_{x_\Spec} \Qprob_{\frac{\eparam+\eparam'}{2}}(x_\Spec)},
\end{eqnarray*}
with an analogous inequality for the term involving
$\Qprob_{\eparam'}$.
Consequently, we have the upper bound
\begin{eqnarray}
\label{EqnBouOne}
\Exs_{\frac{\eparam + \eparam'}{2}}
\big[\weirdkull{\eparam[X_\Spec]}{\eparam'[X_\Spec]} \big] & \leq &
\log \frac{\big[\sum_{x_\Spec} \Qprob_{\eparam}(x_\Spec) \big] \;
\big[\sum_{x_\Spec} \Qprob_{\eparam'}(x_\Spec)\big]}
{\big[\sum_{x_\Spec}
\Qprob_{\frac{\eparam+\eparam'}{2}}(x_\Spec)\big]^2},
\end{eqnarray}
which we exploit momentarily.

In order to use this bound, let us upper bound the quantity
\begin{eqnarray*}
\Delta(\eparam, \eparam') & \defn & \Exs_{\eparam}
\big[\kull{\eparam[X_\Spec]}{(\frac{\eparam+\eparam'}{2})[X_\Spec]}
\big] + \Exs_{\eparam'}
\big[\kull{\eparam'[X_\Spec]}{(\frac{\eparam+\eparam'}{2})[X_\Spec]}
\big].
\end{eqnarray*}
By the definition of the Kullback-Leibler divergence, we have
\begin{multline}
\label{EqnBouTwo}
\Delta(\eparam, \eparam') = \sum_{u \in \Neigh(s) \backslash t}
(\meanpar_{su} - \meanpar'_{su}) \, (\eparam_{su} - \eparam'_{su}) +
\sum_{v \in \Neigh(t) \backslash s} (\meanpar_{tv} - \meanpar'_{tv})
\, (\eparam_{tv} - \eparam'_{tv}) \\
+ \Exs_{\eparam} \log
\frac{\Qprob_{\frac{\eparam+\eparam'}{2}}(X_\Spec)}{\Qprob_{\eparam}(X_\Spec)}
+ \Exs_{\eparam'} \log \frac
{\Qprob_{\frac{\eparam+\eparam'}{2}}(X_\Spec)}{\Qprob_{\eparam'}(X_\Spec)}
\end{multline}
In this equation, the quantities $\meanpar$ and $\meanpar'$ denote
mean parameters computed under the distributions $\mprob_\eparam$ and
$\mprob_{\eparam'}$ respectively. But by Jensen's inequality, we have
the upper bound
\begin{eqnarray}
\label{EqnBouThree}
\Exs_{\eparam} \log
\frac{\Qprob_{\frac{\eparam+\eparam'}{2}}(X_\Spec)}{\Qprob_{\eparam}(X_\Spec)}
& \leq & \log \frac{\sum_{x_\Spec}
\Qprob_{\frac{\eparam+\eparam'}{2}}(x_\Spec)}{\sum_{x_\Spec}
\Qprob_\eparam(x_\Spec)},
\end{eqnarray}
with an analogous upper bound for the term involving $\eparam'$.

Combining the bounds~\eqref{EqnBouOne}, ~\eqref{EqnBouTwo}
and~\eqref{EqnBouThree}, we obtain
\begin{eqnarray*}
\Exs_{\frac{\eparam + \eparam'}{2}}
\big[\weirdkull{\eparam[X_\Spec]}{\eparam'[X_\Spec]} \big] & \leq &
\sum_{u \in \Neigh(s) \backslash t}
(\meanpar_{su} - \meanpar'_{su}) \, (\eparam_{su} - \eparam'_{su}) +
\sum_{v \in \Neigh(t) \backslash s} (\meanpar_{tv} - \meanpar'_{tv})
\, (\eparam_{tv} - \eparam'_{tv}).
\end{eqnarray*}
Finally, since $\sum_{u \in \Neigh(s)} |\eparam_{us}| \leq \maxneigh$
by the definition~\eqref{EqnDefnMaxNeigh} (and similarly for the
neighborhood of $t$), we conclude that
\begin{eqnarray*}
\Exs_{\frac{\eparam + \eparam'}{2}}
\big[\weirdkull{\eparam[X_\Spec]}{\eparam'[X_\Spec]} \big] & \leq & 2
\maxneigh \, \max_{u \in \{s,t\}, v \in \vertex} |\meanpar_{uv} -
\meanpar'_{uv}|.
\end{eqnarray*}
Combining this upper bound with the lower bound~\eqref{EqnAve} yields
the claim.

\section{Discussion}
\label{SecDiscuss}

In this paper, we have analyzed the information-theoretic limits of
binary graphical model selection in a high-dimensional framework, in
which the sample size $\numobs$, number of graph vertices $\pdim$,
number of edges $\kdim$ and/or the maximum vertex degree $\degmax$ are
allowed to tend to infinity.  We proved four main results,
corresponding to both necessary and sufficient conditions for the
class $\Gcldeg$ of graphs on $\pdim$ vertices with maximum vertex
degree $\degmax$, as well as for the class $\Gcledge$ of graphs on
$\pdim$ vertices with at most $\kdim$ edges.  More specifically, for
the class $\Gcldeg$, we showed that any algorithm requires at least
$\numobs > c \, \degmax^2 \log \pdim$ samples, and we demonstrated an
algorithm that succeeds using $\numobs < c' \, \degmax^3 \log \pdim$
samples.  Our two main results for the class $\Gcldeg$ have a similar
flavor: we show that any algorithm requires at least $\numobs > c \,
\kdim \log \pdim$ samples, and we demonstrated an algorithm that
succeeds using $\numobs < c' \, \kdim^2 \log \pdim$ samples.  Thus,
for graphs with constant degree $\degmax$ or a constant number of
edges $\kdim$, our bounds provide a characterization of the
information-theoretic complexity of binary graphical selection that is
tight up to constant factors.  For growing degrees or edge numbers,
there remains a minor gap in our conditions.

In terms of open questions, one immediate issue is to close the
current gap between our necessary and sufficient conditions; as
summarized above, these gaps are of order $\degmax$ and $\kdim$ for
$\Gcldeg$ and $\Gcledge$ respectively.  We note that previous work by
Ravikumar et al.~\cite{RavWaiLaf08} has shown that a computationally
tractable method, based on $\ell_1$-regularization and logistic
regression, can recover binary graphical models using $\numobs =
\Omega(\degmax^3 \log \pdim)$ samples.  This result is consistent with
the theory given here, and it would be interesting to determine
whether or not their algorithm, appealing due to its computational
tractability, is actually information-theoretically optimal.
Moreover, in the current paper, although we have focused exclusively
on binary graphical models with pairwise interactions, many of the
techniques and results (e.g., constructing ``packings'' of graph
classes, Fano's lemma and variants, large deviations analysis) applies
to more general classes of discrete graphical models, and it would be
interesting to explore extensions in this direction.

\subsection*{Acknowledgements}
This work was partially supported by NSF grants CAREER-0545862 and
DMS-0528488 to MJW.


\appendix

\section{A separation lemma}
\label{AppLemKLTech}

In this appendix, we prove the following lemma, which plays a key role
in the proofs of both Lemmas~\ref{LemMatch} and~\ref{LemMeanSep}.
Given an edge $e = (s,t)$ and some subset $U \subseteq \vertex
\backslash \{s,t\}$, recall that
$\newweird{e}{x_U}{\eparam}{\eparam'}$ denotes the
divergence~\eqref{EqnWeirdKull} applied to the conditional
distributions of $(X_u, X_v \, \mid X_\Aset = x_\Aset)$, as defined
explicitly in equation~\eqref{EqnDefnSubWeird}.
\begin{lemma} 
\label{LemKLTech}
Consider two distinct graphs $\graph = (\vertex, \edge)$ and $\graph'
= (\vertex, \edge')$, with associated parameter vectors $\eparam$ and
$\eparam'$.  Given an edge $(s,t) \in \edge \backslash \edge'$ and any
subset $U \subseteq \vertex \backslash \{s,t\}$, we have
\begin{eqnarray}
\label{EqnKLTech}
\newweird{e_\ell}{x_U} {\eparam}{\eparam'} & \geq & \frac{1}{3 \exp(2
\maxneigh) + 1} \sinh^2(\frac{\eparam_{uv}}{4}).
\end{eqnarray}
\end{lemma} 
\begin{proof}  To lighten notation, we define
$\epsilon \defn \frac{2}{3 \exp(2 \maxneigh) + 1}
\sinh^2(\frac{\eparam_{uv}}{4}) > 0$.  Note that from the
definition~\eqref{EqnDefnMaxNeigh}, we have $\maxneigh \geq
|\eparam_{uv}|$, which implies that $\epsilon \leq 2$.
For future reference, we also note the relation
\begin{eqnarray}
\label{EqnUseful}
\big[\exp(\frac{\eparam_{uv}}{4}) -
\exp(-\frac{\eparam_{uv}}{4})\big]^2 & = & 2\epsilon + 6 \epsilon
\exp(2 \maxneigh).
\end{eqnarray}

With this set-up, our argument proceeds via proof by contradiction.
In particular, we assume that 
\begin{eqnarray}
\label{EqnContradict}
\newweird{e_\ell}{x_U} {\eparam}{\eparam'} & \leq & \epsilon/2
\end{eqnarray}
and then derive a contradiction.  Recall from
equation~\eqref{EqnQprob} our notation $\Qprob_\eparam(x_\Aset)$ for
the unnormalized distribution applied to the subset of variables
$x_\Aset = \{ x_i, \; i \in \Aset \}$.  With a little bit of algebra,
we find that
\begin{eqnarray*}
\newweird{e_\ell}{x_U} {\eparam}{\eparam'} & = &
\log\frac{\Qprob_\eparam(x_U) \; \Qprob_{\eparam'}(x_U)}
{\biggr(\sum_{\substack{ z_1 \ldots z_\pdim \\ z_{S} = x_{U}}
}\sqrt{\Qprob_\eparam(z) \Qprob_{\eparam'}(z)}\biggr)^2}.
\end{eqnarray*}

Let us introduce some additional shorthand so as to lighten notation
in the remainder of the proof.  First we define $\beta(x) \, \defn \,
\sqrt{\Qprob_\eparam(x) \, \Qprob_{\eparam'}(x)}$, as well as
$\alpha(x) \defn
\sqrt{\frac{\Qprob_{\eparam}(x)}{\Qprob_{\eparam'}(x)}}$.  Now observe
that $\alpha(x) = \exp(\Delta(x)/2)$, where $\Delta(x) \defn
\sum_{(s,t) \in \edge} \eparam_{st} x_s x_t - \sum_{(s,t) \in \edge'}
\eparam'_{st} x_s x_t$.  Observe that Lemma~\ref{LemFlip} in
Appendix~\ref{AppLemFlip} characterizes the behavior of $\Delta(x)$
under changes to $x$.  Finally, we define the set 
\begin{eqnarray*}
\Yset(x_U) & \defn & \big \{ y \in \{-1, +1\}^\pdim \, \mid \; y_i = x_i
\quad \mbox{for all $i \in U$} \big \},
\end{eqnarray*}
corresponding to the subset of configurations $y \in \{-1, +1\}^\pdim$
that agree with $x_U$ over the subset $U$.

From the definitions of $\alpha$ and $\beta$, we observe that
\begin{eqnarray}
\big[\sum_{y \in \Yset(x_U) } \Qprob_\eparam(y)\big] \; \big[\sum_{y
\in \Yset(x_U) } \Qprob_{\eparam'}(y) \big] & = & \big[\sum_{y \in
\Yset(x_U) } \alpha(y) \beta(y) \big] \; \big[\sum_{y \in \Yset(x_U) }
\frac{\alpha(y)}{\beta(y)} \big] \notag \\
\label{EqnQuad}
& \leq & (1+\epsilon)^{2} \big[\sum_{y \in \Yset(x_U)} \beta(y)
\big]^2,
\end{eqnarray}
where the inequality follows from the fact $\epsilon \leq 2$, our original
assumption~\eqref{EqnContradict}, and the elementary relations
\begin{equation*}
\exp(z) \; < \; (1+ 2z) \; < \; (1+2z)^2 \qquad \mbox{for all $z \in
(0,1]$.}
\end{equation*}

Now consider the set of quadratics in $t$, one for each $y \in
\Yset(x_U)$, given by 
\begin{equation*}
\beta(y) \alpha(y) - 2(1+\epsilon) \beta(y) \,
t + \frac{\beta(y)}{\alpha(y)} t^2 \; = \; 0.
\end{equation*}
Summing these quadratic equations over $y \in \Yset(x_U)$ yields
\begin{eqnarray*}
q(t) \; \defn \; \sum_{y \in \Yset(x_U)} \beta(y) \alpha(y) - 2 t
 (1+\epsilon) \sum_{y \in \Yset(x_U)} \beta(y) +t^2 \sum_{y \in
 \Yset(x_U)} \frac{\beta(y)}{\alpha(y)} \; = \; 0,
\end{eqnarray*}
which by which by equation~\eqref{EqnQuad} must have two real roots.

Let $\tmin$ denote the value of $t$ at which $q(\cdot)$ achieves its
minimum.  By the quadratic formula, we have
\begin{eqnarray*}
\tmin & = & \frac{(1+\epsilon) \sum_{y \in \Yset(x_U)} \beta(y)}
{\sum_{y \in \Yset(x_U)} (\beta(y)/\alpha(y))} \; > \; 0.
\end{eqnarray*}
Since $q(\tmin) < 0$, we obtain
\begin{eqnarray}
\label{eq:quadcont}
 2 \epsilon \tmin \sum_{y \in \Yset(x_U)} \beta(y) & > & \sum_{y \in
 \Yset(x_U)} \beta(y) \tmin \big[\sqrt{\tmin/\alpha(y)} -
 \sqrt{\alpha(y)/\tmin} \big]^2.
\end{eqnarray}
Using the notation
\begin{eqnarray*}
\newSet & \defn & \big \{ y \in \Yset(x_U) \; \mid \; \max \{
  \tmin/\alpha(y), \: \alpha(y)/\tmin \} < \exp(\eparam_{uv}/2) \big
  \},
\end{eqnarray*}
we can rewrite equation~\eqref{eq:quadcont} as
\begin{eqnarray}
\label{EqnWorm}
2 \epsilon \tmin \sum_{y \in \newSet} \beta(y) & > & \sum_{y \notin
\newSet} \beta(y) \tmin \biggr \{ \big[\sqrt{\tmin/\alpha(y)} -
\sqrt{\alpha(y)/\tmin} \big]^2 - 2 \epsilon \biggr \} \\
& \stackrel{(a)}{\geq} & 2 \epsilon \tmin \sum_{y \notin \newSet} 3
\beta(y) \; \exp(2 \maxneigh), \nonumber
\end{eqnarray}
where inequality (a) follows from the definition of $\newSet$, the
monotonically increasing nature of the function \mbox{$f(s) =
(s-\frac{1}{s})^{2}$} for $s \geq 1$, and the
relation~\eqref{EqnUseful}.

From Lemma~\ref{LemFlip}, for each $y \in \newSet$, we obtain a
configuration $a \notin \newSet$ by flipping either $u$, $v$ or
both. Note that at most three configurations $y \in \newSet$ can yield
the same configuration $z \notin \newSet$.   Since these flips do not decrease
$\beta(y)$ by more than a factor of $\exp(2 \maxneigh)$.
we conclude that
\begin{eqnarray*}
\sum_{y \in \newSet} \beta(y) & \leq & 3 \, \exp(2 \maxneigh) \;
\sum_{y \notin \newSet} \beta(y),
\end{eqnarray*}
which is a contradiction of equation~\eqref{EqnWorm}. Hence the
quadratic $q(\cdot)$ cannot have two real roots, which contradicts our
initial assumption~\eqref{EqnContradict}. \\

\hfill \end{proof}


\section{Proof of a flipping lemma}
\label{AppLemFlip}

\newcommand{\del}[2]{\ensuremath{\delta_{#1}(#2)}}
\newcommand{\delprime}[2]{\ensuremath{\delta'_{#1}(#2)}}

\newcommand{\deltilde}[2]{\ensuremath{\gamma_{#1}(#2)}}
\newcommand{\delprimetilde}[2]{\ensuremath{\gamma_{#1}(#2)}}
\newcommand{\har}[1]{\ensuremath{\mu(#1)}}
\newcommand{\harprime}[1]{\ensuremath{\mu'(#1)}}

It remains to state and prove a lemma that we exploited in the proof of
Lemma~\ref{LemKLTech} from Appendix~\ref{AppLemKLTech}.
\begin{lemma}
\label{LemFlip}
Consider distinct models $\eparam$ and $\eparam'$, and for each $x \in
\{-1, +1\}^\pdim$, define
\begin{eqnarray}
\Delta(x) & \defn & \sum_{(u,v) \in \edge} \eparam_{uv} x_u x_v -
\sum_{(u,v) \in \edge'} \eparam'_{uv} x_u x_v.
\end{eqnarray}
Then for any edge $(s,t) \in \edge\backslash \edge'$ and for any
configuration $x \in \{-1, +1\}^\pdim$, flipping either $x_s$ or $x_t$
(or both) changes $\Delta(x)$ by at least $|\eparam_{st}|$.
\end{lemma}
\begin{proof}
We use $\Neigh(s)$ and $\Neigh'(s)$ to denote the neighborhood sets of
$s$ in the graphs $\graph = (\vertex, \edge)$ and $\graph' = (\vertex,
\edge')$ respectively, with analogous notation for the sets
$\Neigh(t)$ and $\Neigh'(t)$.  We then define
\begin{equation*}
\del{s}{x} \, \defn \, \sum_{u \in \Neigh(s) \backslash \Neigh'(s)}
\eparam_{su} x_u, \qquad \mbox{and} \qquad \delprime{s}{x} \, \defn \,
\sum_{u \in \Neigh'(s) \backslash \Neigh(s)} \eparam'_{su} x_u,
\end{equation*}
with analogous definitions for the quantities $\del{t}{x}$ and
 $\delprime{t}{x}$.  Similarly, we define
\begin{align*}
\deltilde{s}{x} \defn  \sum_{u \in \Neigh(s) \cap \Neigh'(s)}
\eparam_{su} x_{u}, \qquad \mbox{and} \quad
\delprimetilde{t}{x} \, \defn\, \sum_{v \in \Neigh(t) \cap \Neigh'(t)}
\eparam_{tv } x_{v}.
\end{align*}

Now, let the contribution to the first (respectively second) term of
$\cE$ not involving $s$ and $t$ be $r_{i}$ (respectively $r_{j}$),
namely 
\begin{equation*}
\har{x} \defn \, \sum_{\substack{ (u,v) \in \edge_{i} \\ u,v \notin
 \{s,t\} } } \eparam_{uv} x_u x_v, \qquad \mbox{and} \qquad
 \harprime{x} \, \defn \, \sum_{\substack{(u,v) \in \edge' \\ u,v
 \notin \{s,t\} }} \eparam_{uv} x_u x_v.
\end{equation*}

With this notation, we first proceed via proof by contradiction to
show that $\Delta(x)$ must change when $(x_s, x_t)$ are flipped: to
the contrary, suppose that for $\Delta(x)$ stays fixed for all four
choices $(x_s, x_t) \in \{-1,+1\}^2$.  We then show that this
assumption implies that $\eparam_{st} = 0$.  Note that both of the
terms $\del{s}{x}$ and $\del{t}{x}$ include a contribution from the
edge $(s,t)$.  When $(x_s, x_t) = (+1, +1)$, we have
\begin{equation*}
\label{EqnPlusPlus}
(\del{s}{x} - \eparam_{st}) + (\del{t}{x}-\eparam_{st}) + \eparam_{st}
+ \har{x} + \deltilde{s}{x} + \deltilde{t}{x} = \delprime{s}{x} +
\delprime{t}{x} + \harprime{x} + \delprimetilde{s}{x} + \delprimetilde{t}{x}
+ \Delta(x), \qquad
\end{equation*}
whereas when $(x_s, x_t) = (-1, -1)$, we have
\begin{equation*}
\label{EqnNegNeg}
-(\del{s}{x} + \eparam_{st}) - (\del{t}{x} +
\eparam_{st})+\eparam_{st}+\har{x} - \deltilde{s}{x} - \deltilde{t}{x} =
-\delprime{s}{x} - \delprime{t}{x} + \harprime{x} - \delprime{s}{x} -
\delprimetilde{t}{x} + \Delta(x). \qquad
\end{equation*}
Adding these two equations together yields the equality 
\begin{equation}
\label{EqnEqualOne}
\har{x} - \eparam_{st} = \harprime{x} + \Delta(x).
\end{equation}
On the other hand, for $(x_s, x_t) = (-1,+1)$, we have
\begin{equation*}
-(\del{s}{x} - \eparam_{st}) + (\del{t}{x} +
\eparam_{st})-\eparam_{st}+\har{x} - \deltilde{s}{x} + \deltilde{t}{x} =
-\delprime{s}{x} + \delprime{t}{x} + \harprime{x} - \delprimetilde{s}{x} +
\delprimetilde{t}{x} + \Delta(x),
\end{equation*}
and for $(x_s, x_t) = (+1, -1)$, we have
\begin{equation*}
(\del{s}{x}+\eparam_{st}) -
(\del{t}{x}-\eparam_{st})-\eparam_{st}+\har{x} + \deltilde{s}{x} -
\deltilde{t}{x} = \delprime{s}{x} - \delprime{t}{x} + \harprime{x} +
\delprimetilde{s}{x} - \delprimetilde{t}{x} + \Delta(x).
\end{equation*}
Adding together these two equations yields
\begin{equation}
\label{EqnEqualTwo}
\har{x} + \eparam_{st} = \harprime{x} + \Delta(x).
\end{equation}
Note that equations~\eqref{EqnEqualOne} and~\eqref{EqnEqualOne} cannot
hold simultaneously unless $\eparam_{st} = 0$, which implies that our
initial assumption---namely, that $\Delta(x)$ does not change as we
vary $(x_s, x_t) \in \{-1, -1\}^2$----was false.

Finally, we show that the change in $|\Delta(x)|$ must be at least
$|\eparam_{st}|$.  For each pair $(i,j) \in \{-1, +1\}^2$, let
$\cE_{ij} = \Delta(x \, \mid \, x_{s} = i, x_{t}=j)$ be the value of
$\Delta(x)$ when $x_{s} = i$ and $x_{t} = j$.  Suppose that for some
constant $c$ and $\delta > 0$, we have $\cE_{ij} \in [c - \epsilon, c
+ \epsilon]$ for all $(i,j)$.  By following the same reasoning as
above, we obtain the inequalities $\har{x} - \eparam_{st} \geq
\harprime{x} + c - \epsilon$ and $\har{x} + \eparam_{st} \leq
\harprime{x} + c + \epsilon$, which together imply that $\eparam_{st}
\leq \epsilon$.  In a similar manner, we obtain the inequalities
$\har{x} + \eparam_{st} \geq \harprime{x} + c - \epsilon$ and $\har{x}
- \eparam_{st} \leq \harprime{x} + c + \epsilon$, which imply that
$-\eparam_{st} \leq \epsilon$, thereby completing the proof.

\hfill \end{proof}


\ifthenelse{\equal{\doctype}{TECH}}
{\bibliographystyle{plain} }
{ 
}

\bibliography{mjwain_super}

\begin{thebibliography}{10}

\bibitem{AhmSonXin08}
A.~Ahmedy, L.~Song, and E.~P. Xing.
\newblock Time-varying networks: Recovering temporally rewiring genetic
  networks during the life cycle of drosophila melanogaster.
\newblock Technical Report arXiv, Carngie Mellon University, 2008.

\bibitem{Alon}
N.~Alon and J.~Spencer.
\newblock {\em The {P}robabilistic {M}ethod}.
\newblock Wiley Interscience, New York, 2000.

\bibitem{BanGhaAsp08}
O.~Bannerjee, , L.~El Ghaoui, and A.~d'Aspremont.
\newblock Model selection through sparse maximum likelihood estimation for
  multivariate {G}aussian or binary data.
\newblock {\em Jour. Mach. Lear. Res.}, 9:485--516, March 2008.

\bibitem{Baxter}
R.~J. Baxter.
\newblock {\em Exactly solved models in statistical mechanics}.
\newblock Academic Press, New York, 1982.

\bibitem{Besag86}
J.~Besag.
\newblock On the statistical analysis of dirty pictures.
\newblock {\em Journal of the Royal Statistical Society, Series B},
  48(3):259--279, 1986.

\bibitem{BreMosSly08}
G.~Bresler, E.~Mossel, and A.~Sly.
\newblock Reconstruction of markov random fields from samples: Some easy
  observations and algorithms.
\newblock Technical Report arXiv, UC Berkeley, 2008.

\bibitem{Brown86}
L.D. Brown.
\newblock {\em Fundamentals of statistical exponential families}.
\newblock Institute of Mathematical Statistics, Hayward, CA, 1986.

\bibitem{chickering:95}
D.~Chickering.
\newblock Learning {B}ayesian networks is {NP}-complete.
\newblock {\em Proceedings of AI and Statistics}, 1995.

\bibitem{Chow68}
C.~K. Chow and C.~N. Liu.
\newblock Approximating discrete probability distributions with dependence
  trees.
\newblock {\em IEEE Trans. Info. Theory}, IT-14:462--467, 1968.

\bibitem{Cover}
T.M. Cover and J.A. Thomas.
\newblock {\em Elements of Information Theory}.
\newblock John Wiley and Sons, New York, 1991.

\bibitem{CsiTal06}
I.~Csisz\'{a}r and Z.~Talata.
\newblock Consistent estimation of the basic neighborhood structure of {M}arkov
  random fields.
\newblock {\em The Annals of Statistics}, 34(1):123--145, 2006.

\bibitem{DurbinEtal}
R.~Durbin, S.~Eddy, A.~Krogh, and G.~Mitchison, editors.
\newblock {\em Biological Sequence Analysis}.
\newblock Cambridge University Press, Cambridge, 1998.

\bibitem{FriedHasTib2007}
J.~Friedman, T.~Hastie, and R.~Tibshirani.
\newblock Sparse inverse covariance estimation with the graphical {L}asso.
\newblock {\em Biostatistics}, 2007.

\bibitem{Geman84}
S.~Geman and D.~Geman.
\newblock Stochastic relaxation, {Gibbs} distributions, and the {Bayesian}
  restoration of images.
\newblock {\em {IEEE} Trans. {PAMI}}, 6:721--741, 1984.

\bibitem{Hasminskii78}
R.~Z. Has'minskii.
\newblock A lower bound on the risks of nonparametric estimates of densities in
  the uniform metric.
\newblock {\em Theory Prob. Appl.}, 23:794--798, 1978.

\bibitem{Hoeffding63}
W.~Hoeffding.
\newblock Probability inequalities for sums of bounded random variables.
\newblock {\em Journal of the American Statistical Association}, 58:13--30,
  1963.

\bibitem{IbrHas81}
I.~A. Ibragimov and R.~Z. Has'minskii.
\newblock {\em Statistical {E}stimation: {A}symptotic {T}heory}.
\newblock Springer-Verlag, New York, 1981.

\bibitem{Ising25}
E.~Ising.
\newblock Beitrag zur theorie der ferromagnetismus.
\newblock {\em Zeitschrift f\"{u}r Physik}, 31:253--258, 1925.

\bibitem{JiSey96}
C.~Ji and L.~Seymour.
\newblock A consistent model selection procedure for {M}arkov random fields
  based on penalized pseudolikelihood.
\newblock {\em Annals of {A}pplied {P}rob.}, 6(2):423--443, 1996.

\bibitem{KalBuh07}
M.~Kalisch and P.~B\"{u}hlmann.
\newblock Estimating high-dimensional directed acyclic graphs with the {PC}
  algorithm.
\newblock {\em Journal of Machine Learning Research}, 8:613--636, 2007.

\bibitem{Meinshausen06}
N.~Meinshausen and P.~B\"uhlmann.
\newblock High-dimensional graphs and variable selection with the {L}asso.
\newblock {\em Annals of Statistics}, 34:1436--1462, 2006.

\bibitem{RavWaiLaf08}
P.~Ravikumar, M.~J. Wainwright, and J.~Lafferty.
\newblock High-dimensional graph selection using $\ell_1$-regularized logistic
  regression.
\newblock {\em Annals of Statistics}, 2008.
\newblock To appear.

\bibitem{Ravetal08}
P.~Ravikumar, M.~J. Wainwright, G.~Raskutti, and B.~Yu.
\newblock High-dimensional covariance estimation: Convergence rates of
  $\ell_1$-regularized log-determinant divergence.
\newblock Technical report, Department of Statistics, UC Berkeley, September
  2008.

\bibitem{Rot09}
A.~J. Rothman, P.~J. Bickel, E.~Levina, and J.~Zhu.
\newblock Sparse permutation invariant covariance estimation.
\newblock {\em Electronic Journal of Statistics}, 2009.

\bibitem{SDM09:itw}
N.~Santhanam, J.~Dingel, and O.~Milenkovic.
\newblock On modeling gene regulatory networks using markov random fields.
\newblock In {\em Information Theory Workshop}, Volos, Greece, June 2009.

\bibitem{spirtes:00}
P.~Spirtes, C.~Glymour, and R.~Scheines.
\newblock Causation, prediction and search.
\newblock {\em MIT Press}, 2000.

\bibitem{Veg07}
F.~Vega-{R}edondo.
\newblock {\em Complex social networks}.
\newblock Econometric Society Monographs. Cambridge University Press,
  Cambridge, 2007.

\bibitem{WaiJor08}
M.~J. Wainwright and M.~I. Jordan.
\newblock Graphical models, exponential families and variational inference.
\newblock {\em Foundations and Trends in Machine Learning}, 1(1--2):1---305,
  December 2008.

\bibitem{WasFau}
S.~Wasserman and K.~Faust.
\newblock {\em Social network analysis: {M}ethods and applications}.
\newblock Cambridge University Press, New York, NY, 1994.

\bibitem{YanBar99}
Y.~Yang and A.~Barron.
\newblock Information-theoretic determination of minimax rates of convergence.
\newblock {\em Annals of Statistics}, 27(5):1564--1599, 1999.

\bibitem{Yu97}
B.~Yu.
\newblock Assouad, {F}ano and {L}e {C}am.
\newblock In {\em Festschrift for {L}ucien {L}e {C}am}, pages 423--435.
  Springer-Verlag, Berlin, 1997.

\bibitem{YuaLin07}
M.~Yuan and Y.~Lin.
\newblock Model selection and estimation in the {G}aussian graphical model.
\newblock {\em Biometrika}, 94(1):19--35, 2007.

\end{thebibliography}

\end{document}